\documentclass[11pt]{article}
\usepackage{amsmath}
\usepackage{natbib}
\usepackage{latexsym, epsfig, amssymb, amsmath, amsthm, graphicx, mathrsfs, booktabs}
\usepackage{rotating, tabularx, setspace,paralist}
\usepackage{bm}
\usepackage{multirow,multicol}
\usepackage{color}
\usepackage{graphicx}
\usepackage[OT1]{fontenc}
\usepackage{hypernat}
\usepackage{lscape}
\usepackage{amsthm,ctable}
\usepackage{anysize}
\renewcommand{\baselinestretch}{1.5}
\marginsize{1.2in}{1in}{0.55in}{1.55in} %

\makeatletter
\def\singlespace{\def\baselinestretch{1}\@normalsize}

\newtheorem{defi}{\sc Definition}
\newtheorem{theo}{\sc Theorem}
\newtheorem{lemma}{\sc Lemma}
\renewcommand{\theequation}{
\arabic{equation}%
}

\makeatother

\begin{document}

\title{Varying Coefficient Panel Data Model with Interactive Fixed Effects%
\date{}
\author{Sanying Feng$^{1}$, Gaorong Li$^{2}$, Heng Peng$^{3}$,   Tiejun Tong$^{3}$
\medskip\\
{\small $^1$School of Mathematics and Statistics, Zhengzhou University, Zhengzhou 450001, China} \\
{\small $^2$Beijing Institute for Scientific and Engineering Computing, Beijing University of} \\
{\small Technology, Beijing 100124,  China}\\
{\small $^3$Department of Mathematics, Hong Kong Baptist University, Hong Kong}
} %
}

\maketitle

\begin{abstract}
In this paper, we propose a varying coefficient panel data model with
unobservable multiple interactive fixed effects that are correlated with
the regressors. We approximate each coefficient function by
B-spline, and propose a robust nonlinear iteration scheme based on
the least squares method to estimate the coefficient functions of interest.  We also establish the asymptotic
theory  of the resulting estimators under certain regularity
assumptions, including the consistency, the convergence rate
and the asymptotic distribution. Furthermore, we develop a least squares dummy variable  
method to study an important special case  of the proposed model: the varying coefficient panel data model with additive fixed effects.
To construct the pointwise confidence intervals
for the coefficient functions, a residual-based  block bootstrap method is proposed to reduce the computational burden as well as to avoid the accumulative errors. Simulation studies and a real data analysis are also carried out to
assess the performance of our proposed methods.

\vskip10pt


\end{abstract}

\section{Introduction}\label{sec1}

Panel data models typically incorporate individual and time effects
to control the heterogeneity in the cross-section and  across the
time-periods. Panel data analysis has attracted considerable
attention in the literature. The methodology for parametric
panel data analysis is quite mature, see, for example, \cite{Arellano2003},  \cite{Hsiao2003}, \cite{Baltagi2005} and the references therein. The individual
and time effects may enter the model additively, or they can
interact multiplicatively that leads to  the so-called
interactive effects or a factor structure.
Panel data models with interactive fixed effects are useful
modelling paradigm. In macroeconomics, incorporating the interactive
effects can account for the heterogenous impact of unobservable
common shocks, while the regressors can be input such  as labor and
capital. Panel data models with interactive fixed effects are used to
incorporate unmeasured skills or unobservable characteristics, or to
study the individual wage rate (see  details in \cite{SuChen2013}). In finance, a combination of unobserved factors and observed
covariates can explain the excess returns of assets.  \cite{Bai2009}
considered the linear panel data  model with interactive
fixed effects:
\begin{eqnarray}\label{LM}
Y_{it}=X_{it}^{\tau}\bm\beta+\lambda_i^{\tau}F_t+\varepsilon_{it},
\quad i=1,\ldots,N,\quad t=1,\ldots,T,
\end{eqnarray}
where $X_{it}$ is a $p\times1$ vector of observable regressors,
$\bm\beta$ is a $p\times1$ vector of unknown coefficients,
$\lambda_i$ is an $r\times1$ vector of factor loadings, $F_t$ is an
$r\times1$ vector of common factors so that
$\lambda_i^{\tau}F_t=\lambda_{i1}F_{1t}+\cdots+\lambda_{ir}F_{rt}$, and
 $\varepsilon_{it}$ are idiosyncratic errors.
In this model, $\lambda_i,F_t$ and $\varepsilon_{it}$ are
all unobserved. We also assume that the dimension $r$ of the factor loadings does not depend on the cross section size $N$
or the time series length $T$.

  A number of
researchers  have developed statistical methods to study  panel data models with interactive fixed effects.
For example, \cite{Holtz1988} estimated
model (\ref{LM}) by quasi-differencing and using lagged variables as
instruments. Their approach, however, ruled out time constant
regressors. \cite{Coakley2002} studied
model (\ref{LM}) by augmenting the regression of $Y$ on $X$ with the
principal components of the ordinary least squares residuals. \cite{Pesaran2006} showed that
the method of \cite{Coakley2002} is inconsistent unless the correlation between $X_{it}$
and $\lambda_i$ tends to be uncorrelated or fully correlated as $N$ tends to infinity. As an
alternative, \cite{Pesaran2006} developed a correlated common effects (CCE)
estimator, in which  model (\ref{LM}) is  augmented by the
cross-sectional averages of $X_{it}$. Although  Pesaran's estimator is consistent,  it does not allow for time-invariant individual regressors.
   \cite{Ahn2001} developed a generalized method of moments
(GMM) estimator for model (\ref{LM}). Their estimator is more
efficient than the least squares estimator under a fixed $T$. However, the
identification of their estimator requires that $X_{it}$ is
correlated with $\lambda_i$, and it is impossible to make testing for the interactive random
effects assumption. \cite{Bai2009} studied the identification, consistency, and
 limiting distribution of the principal component analysis (PCA) estimators
and demonstrated that these estimators are $\sqrt{NT}$ consistent.
\cite{BaiLi2014} investigated the maximum likelihood estimation of model (\ref{LM}).
\cite{WuLi2014} conducted several tests for the existence of individual effects and time effects of model (\ref{LM}).
\cite{Li2016} studied the estimation and inference of common structural breaks in panel
data models with interactive fixed effects using Lasso-type methods. More studies can be found in  \cite{MoonWeidner2010},
\cite{Lee2012}, \cite{SuChen2013}, \cite{MoonWeidner2015},
\cite{LuSu2016}, and many others.

Note that  the aforementioned papers have  focused  on the linear specification of regression relationship in panel
data models with interactive fixed effects.
A natural  extension of model (\ref{LM}) is to consider the  varying
coefficient panel data model with interactive fixed effects:
\begin{eqnarray}\label{VCM}
Y_{it}=X_{it}^{\tau}\bm\beta(U_{it})+\lambda_i^{\tau}F_t+\varepsilon_{it},
\quad i=1,\ldots,N,\quad t=1,\ldots,T,
\end{eqnarray}
where  $\bm\beta(\cdot)$ is a $p\times1$ vector of unknown coefficient functions
 to be estimated. We allow for $\{X_{it}\}$ and/or $\{U_{it}\}$ to
be correlated with $\{\lambda_{i}\}$ alone or with $\{F_t\}$ alone,
or simultaneously correlated with  $\{\lambda_{i}\}$ and
$\{F_t\}$, or correlated with an unknown correlation structure. In fact, $X_{it}$
can be a nonlinear function of $\lambda_i$ and $F_t$. Hence, model
(\ref{VCM}) is a fixed effects model, and assumes an interactive fixed effects linear model
for each fixed time $t$ but allows the coefficients to vary with the
covariate $U_{it}$. This model is attractive because it has an intuitive
interpretation, meanwhile it  retains the unobservable multiple interactive fixed
effects, the general nonparametric characteristics, and the
explanatory power of the linear panel data model.


Model  (\ref{VCM}) is fairly general and it encompasses various panel data models as special
cases. If $X_{it}\equiv 1$ and $p=1$,  model (\ref{VCM}) reduces to the nonparametric
panel data model with interactive fixed effects, which has received much attention in recent years.
\cite{Huang2013} studied the local linear estimation of nonparametric
panel data models with interactive fixed effects.
\cite{SuJin2012} extended the CCE method of \cite{Pesaran2006}
from a linear model to a nonparametric model via the method of sieves. \cite{JinSu2013}
constructed a nonparametric test for poolability in nonparametric regression models with interactive fixed effects.
\cite{Su2015} proposed a consistent nonparametric test for the linearity in large dimensional
panel data model with interactive fixed effects.

If $r=1$ and
$F_t\equiv1$,  model (\ref{VCM}) reduces to the fixed
individual effects panel data varying coefficient model:
$$Y_{it}=X_{it}^{\tau}\bm\beta(U_{it})+\lambda_i+\varepsilon_{it}.$$ This model  has also been widely  studied in the literature. For example,
\cite{Sun2009} considered the estimation  using the local linear regression and the kernel-based weights. \cite{Li2011} considered a nonparametric time varying coefficient model with
fixed effects under the assumption of cross-sectional independence, and proposed
two methods to estimate the trend function and the coefficient functions.
\cite{RodriguezSoberon2014} proposed a new technique to estimate the varying
coefficient functions based on the first-order differences and the local linear regression.
\cite{RodriguezSoberon2015} investigated the model by using the mean transformation technique and the local linear regression.
\cite{LiLian2015} considered the  variable selection for the model using the basis function approximations and the group nonconcave penalized functions.
\cite{Malikov2016} considered the problem of varying coefficient panel data model in the presence of endogenous selectivity and fixed effects.  In addition, if
$\lambda_i\equiv0$  or  $F_t\equiv0$,  model (\ref{VCM}) reduces to the varying
coefficient model with panel data. For the development of  this model, one may refer to, for example,   \cite{Chiang2001}, \cite{Huang2002}, \cite{Huang2004},  \cite{XueZhu2007},  \cite{Cai2007}, \cite{CaiLi2008}, \cite{Wang2008}, \cite{WangXia2009},  and \cite{NohPark2010}.
We note, however, that  most of these papers were 
dealing with the ``large $N$ small $T$" setting.

Despite the rich literature
 in panel data models with interactive fixed effects,
to the best of our knowledge, there is little work on the varying coefficient
panel data models with interactive fixed effects.
Inspired by this,  the main goals of this paper are to estimate the coefficient
functions
$\bm\beta(\cdot)=(\beta_1(\cdot),\ldots,\beta_p(\cdot))^{\tau}$ and
to establish the  asymptotic theory  for the varying
coefficient panel data models with interactive fixed effects when
both  $N$ and  $T$ tend to infinity. To achieve these goals,  we first apply the   B-spline expansion to
estimate the smooth functions in model (\ref{VCM}) due to  its
simplicity.  We then  introduce a novel iterative least squares  procedure
to estimate the coefficient functions and the factor loadings,
and  derive  some asymptotic properties for the proposed  estimators.
Nevertheless, the existence of
the unobservable interactive fixed effects in the model will make the estimation procedure and  the asymptotic theory much  more complicated than those in \cite{Huang2002}.
To apply  the  asymptotic normality for constructing  the pointwise confidence intervals for the coefficient functions, we need  some consistent estimators of the asymptotic variances.  To reduce the  computational burden and to avoid the accumulative errors,
we propose a residual-based block bootstrap procedure to construct the pointwise confidence intervals of the coefficient functions.
Numerical results in Section \ref{sec5} confirm that our proposed estimation procedure works well in a wide range of settings.


The remainder of the  paper is organized as follows. In Section \ref{sec2},  we
propose an estimation procedure for the coefficient functions and
provide a robust iteration algorithm under the identification
restrictions. In Section \ref{sec3}, we establish  the asymptotic
theory of the resulting estimators under some  regularity assumptions as both  $N$ and  $T$ tend to infinity. In
Section \ref{sec4}, as an important special case, we study the
varying coefficient panel data model with additive fixed effects.  To solve it,  we propose a least squares dummy variable  method to estimate the coefficient functions, that avoids the estimation of  the unobserved fixed effects.
In Section \ref{secboot}, a residual-based block bootstrap procedure is developed to
construct the pointwise confidence intervals for the coefficient functions.
In Section \ref{sec5}, a data-driven procedure is proposed to choose the
smoothing parameters, and  numerical studies are
carried out to demonstrate  the efficacy of our proposed methods.
In Section \ref{sec7}, a real dataset is analyzed to augment the derived theoretical results.
Finally, we conclude the paper in Section \ref{sec8} with some remarks. Technical details are given in the Appendices.

\section{Methodology}\label{sec2}

To estimate the coefficient functions $\beta_k(\cdot)$ for $1\le
k\le p$, we consider the widely  used B-spline approximations. Let
$B_k(u)=(B_{k1}(u),\ldots,B_{kL_k}(u))^{\tau}$ be the $(m+1)$th order B-spline basis
functions, where $L_k=l_k+m+1$ is the number of basis
functions in approximating  $\beta_k(u)$,
 $l_k$ is the number of interior knots for
$\beta_k(\cdot)$, and $m$ is the degree of the spline. The interior knots of
the splines can be either equally spaced or placed on the sample
number of observations between any two adjacent knots. With the above basis functions, the coefficient functions $\beta_k(u) $ can be
approximated by
\begin{equation}\label{appr}
\beta_k(u)\approx \sum_{l=1}^{L_k}\gamma_{kl}B_{kl}(u),\quad
k=1,\ldots,p,
\end{equation}
where $\gamma_{kl}$ are the coefficients, and $L_k$ represent the smoothing parameters and they will be selected by  the ``leave-one-subject-out" cross validation. 

Substituting
(\ref{appr}) into model (\ref{VCM}), we have the following approximation:
\begin{equation}\label{FVCM1}
Y_{it}\approx
\sum_{k=1}^{p}\sum_{l=1}^{L_k}\gamma_{kl}X_{it,k}B_{kl}(U_{it})+\lambda_{i}^{\tau}F_{t}+\varepsilon_{it},\quad
i=1,\ldots,N,\quad t=1,\ldots,T.
\end{equation}
Model (\ref{FVCM1}) is a standard  linear regression model with the
interactive fixed effects. As each coefficient function
$\beta_{k}(u)$ in model (\ref{VCM}) is characterized by 
$\bm\gamma_k=(\gamma_{k1},\ldots,\gamma_{kL_{k}})^{\tau}$,  model (\ref{FVCM1}) cannot be estimated
directly due to the unobservable multiple interactive fixed effects
term. In what follows, we propose a robust nonlinear iteration scheme based on the least squares method to estimate the coefficient functions and  to deal with those
 fixed effects.

 For the sake of  convenience, we use
vectors and matrices to present the model  and perform the analysis. Let
\begin{eqnarray*}
{\bm Y}_i=\left(
            \begin{array}{c}
              Y_{i1} \\
              Y_{i2} \\
              \vdots \\
              Y_{iT} \\
            \end{array}
          \right),\quad \bm F=\left(
                                                        \begin{array}{c}
                                                          F_1^{\tau} \\
                                                          F_2^{\tau} \\
                                                          \vdots \\
                                                          F_T^{\tau} \\
                                                        \end{array}
                                                      \right),\quad {\bm \varepsilon}_i=\left(
            \begin{array}{c}
              \varepsilon_{i1} \\
              \varepsilon_{i2} \\
              \vdots \\
              \varepsilon_{iT} \\
            \end{array}
          \right),
\end{eqnarray*}
and
$\Lambda=(\lambda_1,\lambda_2,\ldots,\lambda_N)^{\tau}$ be an
$N\times r$ matrix.  We also define
\begin{eqnarray*}
{\bm B}(u)=\left(
       \begin{array}{ccccccccc}
         B_{11}(u) & \cdots & B_{1L_{1}}(u) & 0 & \cdots & 0 & 0          & \cdots & 0 \\
                   & \vdots &               &   & \vdots &   &            & \vdots &   \\
         0         & \cdots & 0             & 0 & \cdots & 0 & B_{p1}(u)  & \cdots & B_{pL_{p}}(u) \\
       \end{array}
     \right),
\end{eqnarray*}
$R_{it}=(X_{it}^{\tau}{\bm B}(U_{it}))^{\tau},$ and $ {\bm
R}_{i}=(R_{i1},\ldots,R_{iT})^{\tau}$. Let also
$\bm\gamma=(\bm\gamma_1^{\tau},\ldots,\bm\gamma_p^{\tau})^{\tau}$,
where $\bm\gamma_k=(\gamma_{k1},\ldots,\gamma_{kL_{k}})^{\tau}$.
With the above notations, model (\ref{FVCM1}) can  be rewritten as
\begin{eqnarray*}\label{FVCM1-RE}
{\bm Y}_i\approx{\bm R}_i\bm\gamma+{\bm F}\lambda_i+{\bm
\varepsilon}_i,\quad i=1,\ldots,N.
\end{eqnarray*}

Due to potential correlations between the unobservable effects and the
regressors, we treat $F_{t}$ and $\lambda_{i}$ as the fixed effects parameters to be estimated.
To ensure the identifiability of the coefficient functions
$\bm\beta(\cdot)=(\beta_1(\cdot),\ldots,\beta_p(\cdot))^{\tau}$, we  follow \cite{Bai2009} and  impose the
following identification restrictions:
\begin{eqnarray}\label{iden-cond}
{\bm F}^{\tau}{\bm F}/T=I_r \quad {\hbox{and}}\quad
\Lambda^{\tau}\Lambda={\rm diagonal}.
\end{eqnarray}
These two  restrictions can uniquely determine $\Lambda$ and
${\bm F}$. We  then define the objective function as
\begin{equation}\label{LS-Objective}
Q(\bm\gamma,\bm F,\Lambda)=\sum_{i=1}^{N}({\bm Y}_i-{\bm
R}_i\bm\gamma-{\bm F}\lambda_i)^{\tau}({\bm Y}_i-{\bm
R}_i\bm\gamma-{\bm F}\lambda_i)
\end{equation}
subject to the constraint (\ref{iden-cond}).
 Taking partial
derivatives of (\ref{LS-Objective}) with respect to $\lambda_i$ and
setting them equal to zero, we have
\begin{eqnarray}\label{equa-mu}
\widetilde{\lambda}_i=({\bm F}^{\tau}{\bm F})^{-1}{\bm
F}^{\tau}({\bm Y}_i-{\bm R}_i\bm\gamma)=T^{-1}{\bm F}^{\tau}({\bm
Y}_i-{\bm R}_i\bm\gamma).
\end{eqnarray}
Replacing $\lambda_i$ into (\ref{LS-Objective}) by (\ref{equa-mu}), we
have
\begin{eqnarray*}\nonumber
Q(\bm\gamma,\bm F)&=&\sum_{i=1}^{N}({\bm Y}_i-{\bm
R}_i\bm\gamma-{\bm F}\widetilde{\lambda}_i)^{\tau}({\bm
Y}_i-{\bm R}_i\bm\gamma-{\bm F}\widetilde{\lambda}_i)\\
&=&\sum_{i=1}^{N}( {\bm Y}_i-{\bm R}_i\bm\gamma)^{\tau}M_{\bm
F}({\bm Y}_i-{\bm R}_i\bm\gamma),\label{MMF}
\end{eqnarray*}
where
$\widetilde{\Lambda}=(\widetilde{\lambda}_1,\widetilde{\lambda}_2,\ldots,\widetilde{\lambda}_N)^{\tau}$,
and $M_{\bm F}=I_T-{\bm F}({\bm F}^{\tau}{\bm F})^{-1}{\bm
F}^{\tau}=I_T-{\bm F}{\bm F}^{\tau}/T$ is a projection matrix. For
each given  $\bm F$, if $\sum_{i=1}^N{\bm R}_i^{\tau}M_{\bm F}{\bm
R}_i$ is invertible,  the  least squares estimator of $\bm\gamma$
can be uniquely obtained by minimizing  $Q(\bm\gamma,\bm F)$ as
follows:
\begin{eqnarray}\label{LSE}
\hat{\bm\gamma}(\bm F)=\Big(\sum_{i=1}^N{\bm R}_i^{\tau}M_{\bm
F}{\bm R}_i\Big)^{-1}\sum_{i=1}^N{\bm R}_i^{\tau}M_{\bm F}{\bm Y}_i.
\end{eqnarray}
Since the least squares estimator (\ref{LSE}) of $\bm\gamma$ depends
on the unknown common factors $\bm F$, the final solution of
$\bm\gamma$ can be obtained by iteration between $\bm\gamma$ and
$\bm F$ using the following nonlinear equations:
\begin{eqnarray}\label{equation-1}
\hat{\bm\gamma}&=&\Big(\sum_{i=1}^N{\bm R}_i^{\tau}M_{\hat{\bm
F}}{\bm R}_i\Big)^{-1}\sum_{i=1}^N{\bm R}_i^{\tau}M_{\hat{\bm
F}}{\bm Y}_i,\\\label{equation-2} \hat{\bm
F}V_{NT}&=&\left[\frac{1}{NT}\sum_{i=1}^{N}({\bm Y}_i-{\bm
R}_i\hat{\bm\gamma})({\bm Y}_i-{\bm
R}_i\hat{\bm\gamma})^{\tau}\right]\hat{\bm F},
\end{eqnarray}
where $V_{NT}$ is a diagonal matrix consisting of the $r$ largest
eigenvalues of the matrix $(NT)^{-1}\sum_{i=1}^{N}({\bm Y}_i-{\bm
R}_i\hat{\bm\gamma})({\bm Y}_i-{\bm
R}_i\hat{\bm\gamma})^{\tau}$ arranged in decreasing order.
 As noted by \cite{Bai2009}, the iterated solution is somewhat
sensitive to the initial values. \cite{Bai2009} proposed using either the
least squares estimator of $\bm\gamma$ or the principal components
estimate of $\bm F$ to start with. From the numerical studies in Section \ref{sec5}, we find that  the procedure is
more robust when
the principal components estimator of $\bm F$ is used as the initial values. Generally, the poor initial values will result in an exceptionally large number of iterations.
 By (\ref{equa-mu}), (\ref{equation-1}) and
(\ref{equation-2}), we have
\begin{eqnarray}\nonumber
\hat{\Lambda}&=&(\hat{\lambda}_1,\hat{\lambda}_2,\ldots,\hat{\lambda}_N)^{\tau}\\\label{est-lambda}
&=&T^{-1}\Big(\hat{{\bm F}}^{\tau}({\bm Y}_1-{\bm
R}_1\hat{\bm\gamma}), \hat{{\bm F}}^{\tau}({\bm Y}_2-{\bm
R}_2\hat{\bm\gamma}),\ldots,\hat{{\bm F}}^{\tau}({\bm Y}_N-{\bm
R}_N\hat{\bm\gamma})\Big)^{\tau}.~~
\end{eqnarray}

Once we obtain the estimator
$\hat{\bm\gamma}=(\hat{\bm\gamma}_1^{\tau},\ldots,\hat{\bm\gamma}_p^{\tau})^{\tau}$
of $\bm\gamma$  with
$\hat{\bm\gamma}_k=(\hat{\gamma}_{k1},\ldots,\hat{\gamma}_{kL_{k}})^{\tau}$
for $k=1,\ldots,p$, we can estimate $\beta_{k}(u)$ subsequently
by
\begin{eqnarray*}\label{est-coe}
\hat{\beta}_{k}(u)=\sum_{l=1}^{L_k}\hat{\gamma}_{kl}B_{kl}(u),\quad
k=1,\ldots,p.
\end{eqnarray*}
In what follows, we present a  robust
iteration  algorithm for estimating the parameters  $(\bm\gamma, \bm F,\Lambda)$.

\begin{description}
\item[Step 1.] Obtain an initial estimator $(\hat{\bm F},\hat{\Lambda})$ of
$(\bm F,\Lambda)$.

\item[Step 2.] Given $\hat{\bm F}$ and $\hat{\Lambda}$, compute
\begin{eqnarray*}
\hat{\bm\gamma}(\hat{\bm F},\hat{\Lambda})=\Big(\sum_{i=1}^N{\bm
R}_i^{\tau}{\bm R}_i\Big)^{-1}\sum_{i=1}^N{\bm R}_i^{\tau}({\bm
Y}_i-\hat{\bm F}\hat{\lambda}_i).
\end{eqnarray*}

\item[Step 3.] Given $\hat{\bm\gamma}$, compute $\hat{\bm F}$ according to
(\ref{equation-2}) (multiplied by $\sqrt{T}$ due to the restriction
that ${\bm F}^{\tau}{\bm F}/T=I_r$) and calculate $\hat{\Lambda}$
using formula (\ref{est-lambda}).

\item[Step 4.] Repeat Steps 2 and 3 until $(\hat{\bm\gamma}, \hat{\bm
F},\hat{\Lambda})$ satisfy the given convergence criterion.
\end{description}

\section{Regularity assumptions and asymptotic properties}\label{sec3}

To derive some asymptotic properties of the proposed estimators   
we let $\mathcal{F}\equiv\{{\bm F}: {\bm F}^{\tau}\bm F/T=I_r\}$ and
\begin{eqnarray*}\label{DF}
D(\bm F)=\frac{1}{NT}\sum_{i=1}^{N}{\bm R}_i^{\tau}{ M}_{\bm F}{\bm
R}_i-\frac{1}{T}\Big[\frac{1}{N^2}\sum_{i=1}^N\sum_{j=1}^N{\bm
R}_i^{\tau}{ M}_{\bm F}{\bm R}_ja_{ij}\Big],
\end{eqnarray*}
where
$a_{ij}=\lambda_i^{\tau}(\Lambda^{\tau}\Lambda/N)^{-1}\lambda_j.$ To
obtain the unique estimator of $\bm\gamma$ with probability tending
to one, we require that the first term of $D(\bm F)$ on the
right-hand side is positive definite when ${\bm F}$ is observable.
The presence of the second term is because of the unobservable $\bm F$
and $\Lambda$. The reason for this particular form is the
nonlinearity of the interactive effects (see details in \cite{Bai2009}).

\subsection{Regularity assumptions}

In this section, we
 introduce a definition and  present some regularity
assumptions for establishing the asymptotic theory of the resulting estimators.

{\defi\label{defi1} Let ${\mathcal{H}}_d$ define the collection of
all functions on the support $\mathcal{U}$ whose $m$th order derivative
satisfies the H\"{o}lder condition of order $\nu$ with $d\equiv
m+\nu$, where $0<\nu\le1$. That is, for each $h\in{\mathcal{H}}_d$,
there exists a constant $M_0\in(0,\infty)$ such that
$|h^{(m)}(u)-h^{(m)}(v)|\le M_0|u-v|^{\nu}$, for any $u,v\in
\mathcal{U}$.}

\begin{enumerate}
\item[(A1)] The random variable $X_{it}$ is independent and
identically distributed (i.i.d.) cross the $N$ individuals, and there
exists a positive $M$ such that $|X_{it, k}|\le M<\infty$
 for all $k=1, \ldots, p$. The eigenvalues $\rho_1(u)\le \cdots\le\rho_p(u)$ of
$\Omega(u)=E(X_{it}X_{it}^{\tau}|U_{it}=u)$ are bounded away from 0
and $\infty$ uniformly over $u\in \mathcal{U}$, that is, there exist
positive constants $\rho_0$ and $\rho^{*}$ such that
$0<\rho_0\le\rho_1(u)\le \cdots\le\rho_p(u)\le\rho^{*}<\infty$ for
$u\in\mathcal{U}$.

\item[(A2)] The observation variables
$U_{it}$ are chosen independently
according to a distribution $F_U$ on the support $\mathcal{U}$.
Moreover, the density function of $U$,  $f_U(u)$, is uniformly
bounded away from 0 and $\infty$, and continuously differentiable
uniformly over $u\in\mathcal{U}$.

\item[(A3)] $\beta_k(u)\in \mathcal{H}_d$ for all $k=1,\ldots,p$.

\item[(A4)] Let $u_{k1},\ldots,u_{kl_k}$ be the interior knots of the $k$th
coefficient function over $u\in\mathcal{U}=[U_0,U_1]$ for $ k=1,\ldots,p$.
Furthermore,  let $u_{k0}=U_0$ and $ u_{k(l_k+1)}=U_1$. There exists a
positive constant $C_0$ such that $$\dfrac{h_k}{\min_{1\le i\le l_k}h_{ki}}\le
C_0\quad \hbox{and} \quad\dfrac{\max_{1\le k\le p}h_{ki}}{\min_{1\le k\le p}h_{ki}}\le C_0,$$ where
$h_{ki}=u_{ki}-u_{k(i-1)}$ and $ h_k=\max\limits_{1\le i\le (l_k+1)}h_{ki}$.

\item[(A5)] Suppose that $\inf\limits_{\bm F\in \mathcal{F}}D(\bm
F)>0.$

\item[(A6)] $E\|F_t\|^4\le M$ and
$\sum_{t=1}^TF_tF_t^{\tau}\big/T\stackrel{P}\longrightarrow\Sigma_{F}>0$
for some $r\times r$ matrix $\Sigma_{F}$, as $T\rightarrow\infty$.

\item[(A7)] $E\|\lambda_i\|^4\le M$ and
$\Lambda^{\tau}\Lambda/N\stackrel{P}\longrightarrow\Sigma_{\Lambda}>0$
for some $r\times r$ matrix $\Sigma_{\Lambda}$, as
$N\rightarrow\infty$.

\item[(A8)] (i)
 Suppose that  $\varepsilon_{it}$
are 
independent of $X_{js}$, $U_{js}$, $\lambda_j$ and $F_{s}$ for all
$i,t, j$ and $s$ with zero mean and $E(|\varepsilon_{it}|^8)\le M$.

(ii) ${\bm \varepsilon}_1,\ldots,{\bm \varepsilon}_N$ are  independent
of each other with
$E({\bm\varepsilon}_i{\bm\varepsilon}_i^{\tau})=\Omega_i$, where ${\bm \varepsilon}_i=(\varepsilon_{i1},\ldots,\varepsilon_{iT})^{\tau}$ and the
smallest and  largest eigenvalues of $\Omega_i$ are bounded
uniformly for all $i$ and $t$.

\item[(A9)] $\limsup\limits_{N,T}(\max\limits_{k}L_k/\min\limits_k L_k)<\infty.$
\end{enumerate}



Assumptions (A1)--(A4) are  mild conditions that can be  validated
in many practical situations. These conditions have been widely
assumed in the context of varying coefficient models with repeated
measurements, such as \cite{Huang2002}, \cite{Huang2004} and \cite{Wang2008}.
Assumption (A5) is an identification condition for $\bm\gamma$. If $D(\bm F)$ is positive
definite, $\bm\gamma$ can be uniquely determined by
(\ref{equation-1}). If $\bm F$ is observable, the identification
condition for $\bm\gamma$ would be that the first term of $D(\bm F)$
on the right hand side is positive definite. The presence of the
second term is because of the unobservable $\bm F$ and $\Lambda$.
Assumptions (A6) and (A7) imply the existence of $r$ factors.
In this paper, whether $F_t$ or $\lambda_{i}$
has zero mean is not crucial since they are treated as parameters to be estimated.
Assumption (A8) is similar to that used in \cite{Bai2009}, in which the serial correlation and heteroskedasticity are allowed.
Assumption (A9)  can also be found in \cite{NohPark2010}, and this condition is used for the system of general basis
functions $B_{kl}$ including orthonormal bases, non-orthonormal bases
and B-spline.

Let $\|a\|_{L_2}=\{\int_{\mathcal{U}}a^2(u){\rm d}u\}^{1/2}$ be the
$L_2$ norm of any square integrable real-valued function $a(u)$ on
$\mathcal{U}$, and let
$\|A\|_{L_2}=\{\sum_{k=1}^{p}\|a\|_{L_2}^2\}^{1/2}$
be the $L_2$ norm of $A(u)=(a_1(u),\ldots,a_p(u))^{\tau}$, where
$a_k(u)$ are real-valued functions on $\mathcal{U}$ (see details in \cite{Huang2002}).
We define $\hat{\beta}_k(\cdot)$ to be a consistent
estimator of $\beta_k(\cdot)$ if
$\lim\limits_{N, T\rightarrow\infty}\|\hat{\beta}_k(\cdot)-{\beta}_k(\cdot)\|_{L_2}=0$
holds in probability. Define $\delta_{NT}=\min[\sqrt{N},\sqrt{T}]$,
and $L_{N}=\max\limits_{1\le k\le p}L_k$, which tend to infinity as $N$ or $T$ tends to
infinity. 
Let
$\mathcal{D}=\{(X_{it},U_{it},\lambda_i,F_t),i=1,\ldots,N,t=1,\ldots,T\}$.
We use $E_{\mathcal{D}}$ and ${\rm Var}_{\mathcal{D}}$ to denote the
expectation and variance conditional on ${\mathcal{D}}$, respectively.

\subsection{Asymptotic properties}


With an appropriate choice of $L_k$ to balance the bias and variance,
our proposed  estimators have the asymptotic properties including the consistency, 
the convergence rate and the asymptotic distribution.

{\theo\label{theo1} Suppose that assumptions
(A1)--(A9) hold. If $\delta_{NT}^{-2}L_{N}\log L_{N}$ $\rightarrow0$ as
$N\rightarrow\infty$ and  $T\rightarrow\infty$ simultaneously, then

\begin{enumerate}
\item[{\rm (i)}] $\hat{\beta}_k(\cdot), k=1,\ldots,p,$ are uniquely defined with
probability tending to one.
\item[{\rm (ii)}] The matrix ${\bm F}^{\tau}\hat{{\bm F}}/T$ is invertible and
$\|P_{\hat{\bm F}}-P_{\bm F}\|\stackrel{P}\longrightarrow0,$
where ``$\stackrel{P}\longrightarrow$''  denotes  the convergence in probability and $P_{A}=A(A^{\tau}A)^{-1}A^{\tau}$ for a given matrix $A$.
\end{enumerate}
}

The proof of Theorem \ref{theo1} is given in Appendix A. Part (i) of Theorem \ref{theo1} implies that, with
probability tending to one,  we can obtain the unique estimators
$\hat{\beta}_k(\cdot)$ for the unknown coefficient functions ${\beta}_k(\cdot)$  under some regularity assumptions, no matter whether there exist  unobservable multiple interactive fixed
effects in model \eqref{VCM}.  Part (ii) of Theorem \ref{theo1} indicates that
the spaces spanned by $\hat{{\bm F}}$ and ${\bm F}$ are asymptotically consistent. This is a key result to   guarantee that the estimators $\hat{\beta}_k(\cdot)$  have  good asymptotic properties including the optimal convergence rate and consistency and asymptotic normality.

{\theo\label{theo2} Suppose that assumptions (A1)-(A9)
hold. If $\delta_{NT}^{-2}L_{N}\log L_{N}$ $\rightarrow0$ as
$N\rightarrow\infty$ and  $T\rightarrow\infty$ simultaneously, then
$$\|\hat{\beta}_k(u)-\beta_k(u)\|_{L_2}^2=O_P\left(\frac{L_N}{NT}+\frac{L_N}{T^2}+\frac{L_N}{N^2}+L_N^{-2d}\right), \quad k=1,\ldots,p.$$
 }
 \par
 Theorem \ref{theo2} gives the convergence rate   of $\hat{\beta}_k(u)$ for all $k=1,\ldots,p$, 
 and hence establishes the consistency of our proposed estimators under the condition   $\delta_{NT}^{-2}L_{N}\log L_{N}$ $\rightarrow0$ as
$N\rightarrow\infty$ and  $T\rightarrow\infty$ simultaneously.  From the proof of Theorem \ref{theo2} in Appendix A,  we note that the first term in the convergence rate $O_P\left(\frac{L_N}{NT}+\frac{L_N}{T^2}+\frac{L_N}{N^2}+L_N^{-2d}\right)$ is caused by the   stochastic error, the second  and  third terms are caused by the estimation error of the fixed effects ${\bm F}$
and the presence of the serial correlation and heteroskedasticity, and the last term is the error due to
the basis approximation. If we take the appropriate relative rate $T/N\rightarrow c>0$ as $N\rightarrow\infty$ and  $T\rightarrow\infty$ simultaneously,
then we have a more accurate convergence rate as
$$\|\hat{\beta}_k(u)-\beta_k(u)\|_{L_2}^2=O_P\left(\frac{L_N}{NT}+L_N^{-2d}\right),\quad k=1,\ldots,p.$$
Furthermore, if we  take $L_N=O((NT)^{1/(2d+1)})$, then
$$\|\hat{\beta}_k(u)-\beta_k(u)\|_{L_2}^2=O_P\left((NT)^{-2d/(2d+1)}\right),\quad k=1,\ldots,p.$$
This leads to the  optimal convergence rate of order $O_P\left((NT)^{-2d/(2d+1)}\right)$ that holds for i.i.d. data in \cite{Stone1982}.


Let
\begin{eqnarray*}
{\bm Z}_i=M_{\bm F}{\bm R}_i-\frac{1}{N}\sum_{j=1}^Na_{ij}M_{\bm
F}{\bm R}_j.
\end{eqnarray*}
By  Appendix A,  under some appropriate relative rate for $T$ and $N$ and some assumptions, we have
\begin{eqnarray*}\label{repres}
\hat{\bm\gamma}-\widetilde{\bm\gamma}=\left(\frac{1}{NT}\sum_{i=1}^N{\bm Z}_i^{\tau}{\bm Z}_i\right)^{-1}\frac{1}{NT}\sum_{i=1}^N{\bm Z}_i^{\tau}{\bm\varepsilon}_i+o_P(1),
\end{eqnarray*}
where $\widetilde{\bm\gamma}$ is defined in (\ref{term}) in Appendix A.
As $N\rightarrow\infty$ and  $T\rightarrow\infty$ simultaneously,  the variance-covariance matrix $\Phi={\rm Var}(\hat{\bm\gamma}|\mathcal{D})$ of $\hat{\bm\gamma}$ conditioning on $\mathcal{D}$ is the limit in probability of
\begin{eqnarray*}
\Phi^*=\left(\sum_{i=1}^N{\bm Z}_i^{\tau}{\bm Z}_i\right)^{-1}\left(\sum_{i=1}^N{\bm Z}_i^{\tau}\Omega_i{\bm Z}_i\right)\left(\sum_{i=1}^N{\bm Z}_i^{\tau}{\bm Z}_i\right)^{-1}.
\end{eqnarray*}
The variance-covariance matrix of $\hat{\bm\beta}(u)$ conditioning on $\mathcal{D}$ is ${\bm B}(u)\Phi{\bm B}(u)^{\tau}$. Let $\varpi_k$ denote the unit vector in $\mathbb{R}^p$ with 1 in the $k$th coordinate and 0 in all other coordinates for $k=1,\ldots,p$.
 Then the conditional variance of $\hat{\beta}_k(u)$ is
\begin{eqnarray*}\label{variance}
{\rm Var}(\hat{\beta}_k(u)|\mathcal{D})=\varpi_k^{\tau}{\rm Var}(\hat{\bm\beta}(u)|\mathcal{D})\varpi_k, \quad k=1,\ldots,p.
\end{eqnarray*}

Let $\check{\bm\beta}(u)=(\check{\beta}_1(u),\ldots,\check{\beta}_p(u))^{\tau}$,
where $\check{\beta}_k(u)=E(\hat{\beta}_k(u))$ is the mean of $\hat{\beta}_k(u)$ conditioning on $\mathcal{D}$. With the proofs in Appendix A,
 we have the following asymptotic results including the asymptotic normality.

{\theo\label{theo3} Suppose that assumptions (A1)-(A9)
hold. If $\delta_{NT}^{-2}L_{N}\log L_{N}$ $\rightarrow0$  and $L_NT/N\rightarrow0$ as
$N\rightarrow\infty$ and  $T\rightarrow\infty$ simultaneously, then
$$\{{\rm Var}(\hat{\bm\beta}(u)|\mathcal{D})\}^{-1/2}(\hat{\bm\beta}(u)-\check{\bm\beta}(u))
\stackrel{L}\longrightarrow N({\rm\bf{0}},I).$$ In particular,  we
have
$$\{{\rm Var}(\hat{\beta}_k(u)|\mathcal{D})\}^{-1/2}(\hat{\beta}_k(u)-\check{\beta}_k(u))
\stackrel{L}\longrightarrow N({0},1), \quad k=1,\ldots,p,$$ where  ``$\stackrel{L}\longrightarrow$'' denotes the convergence in
distribution.}

{\theo\label{theo4} Under the same assumptions as in Theorem \ref{theo3}, if  $L_N^{2d+1
}/NT\rightarrow\infty$ as
$N\rightarrow\infty$ and  $T\rightarrow\infty$ simultaneously, then
$$\sup_{u\in\mathcal{U}}\Big|\{{\rm Var}(\hat{\beta}_k(u)|\mathcal{D})\}^{-1/2}(\check{\beta}_k(u)-\beta_k(u))\Big|
=o_P(1), \quad k=1,\ldots,p.$$ }

For the varying coefficient model \eqref{VCM} with  unobservable multiple interactive fixed
effects,  Theorems \ref{theo3} and \ref{theo4}  establish the asymptotic normality for the estimators $\hat{\beta}_k(\cdot)$ of the coefficient functions ${\beta}_k(\cdot)$ if $\delta_{NT}^{-2}L_{N}\log L_{N}$ $\rightarrow0$  and $L_NT/N\rightarrow0$ as
$N\rightarrow\infty$ and  $T\rightarrow\infty$ simultaneously. Note that the  results in Theorems \ref{theo3} and \ref{theo4} are very similar to the results
in \cite{Huang2004}.  From the proof of
Theorem \ref{theo4}, we can find that the bias terms of the estimators $\hat{\beta}_k(\cdot)$ are asymptotically negligible in comparison with
 the variance terms when $N\rightarrow\infty$ and  $T\rightarrow\infty$ simultaneously. Hence, if we can obtain a consistent estimator $\widehat{{\rm Var}}(\hat{\beta}_k(u)|\mathcal{D})$
of ${\rm Var}(\hat{\beta}_k(u)|\mathcal{D})$, the asymptotic pointwise confidence intervals for $\beta_k(u)$ can be constructed by
$$\hat{\beta}_k(u)\pm z_{\alpha/2}\{\widehat{{\rm Var}}(\hat{\beta}_k(u)|\mathcal{D})\}^{-1/2},\quad k=1,\ldots,p,$$
where $z_{\alpha/2}$ is the  $(1-\alpha/2)$ quantile  of the standard normal distribution.

\section{Additive fixed effects model }\label{sec4}

In this section, we consider an important special case of model (\ref{VCM}). By letting
$
\lambda_i=(\mu_{i},1)^{\tau}$ and $F_{t}=(1,\xi_{t})^{\tau}$, model (\ref{VCM}) reduces to the  varying coefficient panel data model with additive fixed effects:
\begin{eqnarray}\label{VCMad}
Y_{it}=X_{it}^{\tau}\bm\beta(U_{it})+\mu_i+\xi_t+\varepsilon_{it},
\quad i=1,\ldots,N, \quad t=1,\ldots,T.
\end{eqnarray}
Similar to  (\ref{iden-cond}), for the purpose of identification,
we assume that
\begin{eqnarray}\label{iden-condad}
\sum\limits_{i=1}^{N}\mu_{i}=0 \quad {\hbox{and}}\quad
\sum\limits_{t=1}^{T}\xi_{t}=0.
\end{eqnarray}
Invoking (\ref{appr}), we have
\begin{equation}\label{FVCM1ad}
Y_{it}\approx
\sum_{k=1}^{p}\sum_{l=1}^{L_k}\gamma_{kl}X_{it,k}B_{kl}(U_{it})+\mu_i+\xi_t+\varepsilon_{it},\quad
i=1,\ldots,N,\quad t=1,\ldots,T.
\end{equation}

Note that, if we further assume that $\sum_{t=1}^{T}\xi_{t}^{2}=T$, then $\bm\gamma$
can be estimated by  the iteration procedure
 described in Section \ref{sec2}.  However, we need to estimate the fixed effects $F_{t}$ and $\lambda_i$, where $i=1,\ldots,N$ and $ t=1,\ldots,T$. In order to avoid estimating the fixed effects $F_{t}$ and $\lambda_i$,
 we propose to remove the unknown fixed effects
 by  a least squares dummy variable method based on the identification condition \eqref{iden-condad}.  The estimation procedure is described in what follows.

Let ${\bf 1}_{N}$ denote an $N\times 1$ vector with all elements being ones,
${\bm Y}=({\bm Y}_{1}^{\tau},\ldots,{\bm Y}_{N}^{\tau})^{\tau}$, ${\bf R}=({\bm R}_{1}^{\tau},\ldots,{\bm R}_{N}^{\tau})^{\tau}$,
${\bm \varepsilon}=({\bm \varepsilon}_{1}^{\tau},\ldots,{\bm \varepsilon}_{N}^{\tau})^{\tau}$, ${\bm \mu}=(\mu_{2},\ldots,\mu_{N})^{\tau}$ and ${\bm \xi}=(\xi_{2},\ldots,\xi_{T})^{\tau}$.
By the identification condition \eqref{iden-condad}, we have
$${\bf D}=[-{\bf 1}_{N-1} ~ I_{N-1}]^{\tau}\otimes {\bf 1}_{T} \quad\hbox{and}\quad {\bf S}={\bf 1}_{N}\otimes [-{\bf 1}_{T-1}~ I_{T-1}]^{\tau},$$
 where $\otimes$ denotes the Kronecker product.
Then model $(\ref{FVCM1ad})$ can be rewritten as the matrix form:
\begin{eqnarray*}\label{FVCM1ad2}
{\bm Y}\approx {\bf R}{\bm \gamma}+{\bf D}{\bm \mu}+{\bf S}{\bm \xi}+{\bm \varepsilon}.
\end{eqnarray*}
Next,  we  solve the following optimization problem:
\begin{equation}\label{FVCM1ad3}
\min_{\bm\gamma,\bm\mu,\bm\xi}({\bm Y}-{\bf R}\bm\gamma-{\bf
D}{\bm\mu}-{\bf S}{\bm \xi})^{\tau}({\bm Y}-{\bf R}\bm\gamma-{\bf
D}{\bm\mu}-{\bf S}{\bm \xi}).
\end{equation}
Taking partial derivatives of (\ref{FVCM1ad3}) with respect
to $\bm\mu$ and $\bm\xi$, and setting them equal to zero, we have
$${\bf D}^{\tau}({\bm Y}-{\bf R}\bm\gamma-{\bf
D}{\bm\mu}-{\bf S}{\bm \xi})=0,$$
$${\bf S}^{\tau}({\bm Y}-{\bf R}\bm\gamma-{\bf
D}{\bm\mu}-{\bf S}{\bm \xi})=0.$$
By a simple calculation, we can obtain that
\begin{eqnarray*}
\tilde{{\bm \xi}}&=&({\bf S}^{\tau}{\bf S})^{-1}{\bf S}^{\tau}({\bm Y}-{\bf R}\bm\gamma),\\
\tilde{{\bm \mu}}&=&({\bf D}^{\tau}{\bf D})^{-1}{\bf D}^{\tau}\left[{\bm Y}-{\bf R}\bm\gamma-
            {\bf S}({\bf S}^{\tau}{\bf S})^{-1}{\bf S}^{\tau}({\bm Y}-{\bf R}\bm\gamma)\right].
\end{eqnarray*}
Replacing $\bm\mu$ and $\bm\xi$ in (\ref{FVCM1ad3}) by $\tilde{{\bm \mu}}$ and $\tilde{{\bm \xi}}$ respectively, the parameter
$\bm\gamma$ can be estimated by minimizing
$
({\bm Y}-{\bf R}\bm\gamma)^{\tau}{\bf \Gamma}({\bm Y}-{\bf R}\bm\gamma),
$
where ${\bf \Gamma}={\bf H}(I_{NT}-{\bf S}({\bf S}^{\tau}{\bf S})^{-1}{\bf S}^{\tau})$ and
${\bf H}=I_{NT}-{\bf D}({\bf D}^{\tau}{\bf D})^{-1}{\bf D}^{\tau}$.
 Specifically, the least squares estimator of $\bm\gamma$ is
\begin{eqnarray*}\label{FVCM1ad5}
\breve{\bm\gamma}=\left({\bf R}^{\tau}{\bf \Gamma}{\bf R}\right)^{-1}{\bf R}^{\tau}{\bf \Gamma}{\bm Y}.
\end{eqnarray*}
Then with the estimator
$\breve{\bm\gamma}=(\breve{\bm\gamma}_1^{\tau},\ldots,\breve{\bm\gamma}_p^{\tau})^{\tau}$
of $\bm\gamma$, where
$\breve{\bm\gamma}_k=(\breve{\gamma}_{k1},\ldots,\breve{\gamma}_{kL_{k}})^{\tau}$
for $k=1,\ldots,p$, we can estimate $\beta_{k}(u)$
by
\begin{eqnarray*}\label{FVCM1ad6}
\breve{\beta}_{k}(u)=\sum_{l=1}^{L_k}\breve{\gamma}_{kl}B_{kl}(u),\quad
k=1,\ldots,p.
\end{eqnarray*}

{\theo\label{theo5} Suppose that assumptions (A1)-(A4) and (A8)-(A9)
hold. If $\delta_{NT}^{-2}L_{N}\log L_{N}$ $\rightarrow0$ as
$N\rightarrow\infty$ and  $T\rightarrow\infty$ simultaneously, then
$$\|\breve{\beta}_k(u)-\beta_k(u)\|_{L_2}^2=O_P(L_N/NT+L_N^{-2d}), \quad k=1,\ldots,p.$$  }

As it is not needed to  estimate the fixed effects $F_{t}$ and $\lambda_i$ using the least squares dummy variable method,  the second and third terms of the convergence rate in Theorem \ref{theo2} vanish in Theorem \ref{theo5}  for the varying coefficient panel data model \eqref{VCMad} with additive fixed effects.  The estimators achieve the optimal convergence rate
of order $O_P\left((NT)^{-2d/(2d+1)}\right)$ if we
take $L_N=O((NT)^{1/(2d+1)})$.

\section{A residual-based block bootstrap procedure}\label{secboot}

 In theory,  we can construct the pointwise confidence intervals for the coefficient functions $\beta_k(\cdot)$ by  Theorems \ref{theo3} and \ref{theo4}. But doing so,  we need to derive the consistent estimators of the asymptotic variances of the estimators $\hat{\beta}_{k}(\cdot)$ for $ k=1,\ldots,p$.  Nevertheless, as  the asymptotic variances involve the unknown fixed effects ${\bm F}$
and the covariance matrices $\Omega_{i}$ of ${\bm \varepsilon}_{i}$, it is difficult to obtain the consistent and efficient estimators of the asymptotic variances even if the plug-in method is used to estimate the asymptotic variances of $\hat{\beta}_{k}(\cdot)$.
%
In addition, the standard nonparametric bootstrap procedure cannot be applied to construct the pointwise confidence intervals directly because there exist the serial correlations within the group in the varying coefficient panel data model \eqref{VCM}  with interactive fixed effects. They will not only increase the computational burden and cause the accumulative errors, but also make it more difficult to construct the pointwise confidence intervals. To overcome the limitations,
we
hereby propose a residual-based block bootstrap procedure to construct the pointwise confidence intervals for $\beta_k(\cdot)$ with the detailed algorithm as follows.
\begin{description}
  \item[Step 1.] Fit the varying coefficient panel data model (\ref{VCM}) with interactive fixed effects using the proposed methods in Section \ref{sec2}, and estimate
the residuals $\varepsilon_{it}$ by
\begin{eqnarray*}\label{boot1}
\hat{\varepsilon}_{it}=Y_{it}-
\sum_{k=1}^{p}\sum_{l=1}^{L_k}\hat{\gamma}_{kl}X_{it,k}B_{kl}(U_{it})+\hat{\lambda}_{i}^{\tau}\hat{F}_{t},\quad
i=1,\ldots,N,\quad t=1,\ldots,T.
\end{eqnarray*}

\item[Step 2.] Generate the bootstrap residuals $\varepsilon_{it}^{*}$ by $\hat{\varepsilon}_{it}$
using the block bootstrap method by a two-step procedure:  (i) Similar to the average block length in \cite{InoueShintani2006}, the block
length $l$ is chosen  by  $l= cT^{1/3}$ for some $c>0$.
 (ii) 
To generate the bootstrap samples, the blocks can be overlapping or non-overlapping.
According to \cite{Lahiri1999}, there is little difference in the performance for these two methods. We hence adopt
the non-overlapping method for simplicity  and divide the error data into $m=T/l$ blocks. Then for
 the $N\times T$ matrix $\hat{{\bm\varepsilon}}$, we generate the bootstrap samples $N\times T$ matrix ${\bm \varepsilon^{*}}$  by resampling with replacement the $m$ blocks of columns
of $\hat{{\bm\varepsilon}}$.

\item[Step 3.]  We generate the bootstrap sample $Y_{it}^{*}$ by the following model:
\begin{eqnarray*}\label{boot2}
Y_{it}^{*}=
\sum_{k=1}^{p}\sum_{l=1}^{L_k}\hat{\gamma}_{kl}X_{it,k}B_{kl}(U_{it})+\hat{\lambda}_{i}^{\tau}\hat{F}_{t}+\varepsilon_{it}^{*},\quad
i=1,\ldots,N,\quad t=1,\ldots,T,
\end{eqnarray*}
where $\hat{\gamma}_{kl}, \hat{F}_{t}$ and $\hat{\lambda}_{i}$ are the respective estimators of ${\gamma}_{kl}, {F}_{t}$ and ${\lambda}_{i}$  using the estimation procedure in Section 2. Based on the bootstrap sample $\{(Y_{it}^{*}, X_{it}, U_{it}),   i=1,\ldots,N, t=1,\ldots,T\}$,
we calculate the bootstrap estimator $\hat{\bm\beta}^{(b)}(\cdot)$ also by the estimation procedure in Section 2.

\item[Step 4.] Repeat  Steps 2 and  3 for $B$ times  to get a size $B$ bootstrap estimators $\hat{\bm\beta}^{(b)}(u)$, $b = 1, \ldots, B$.  The bootstrap estimator ${\rm Var}^{*}(\hat{\bm\beta}(u)|\mathcal{D})$ of ${\rm Var}(\hat{\bm\beta}(u)|\mathcal{D})$
is taken as the sample variance of $\hat{\bm\beta}^{(b)}(u)$.
Finally, we construct the asymptotic pointwise confidence intervals for $\beta_{k}(u)$ by
$$\hat{\beta}_k(u)\pm z_{\alpha/2}({\rm Var}^{*}(\hat{\beta}_k(u)|\mathcal{D}))^{1/2},\quad k=1,\ldots,p,$$
where $z_{\alpha/2}$ is the $(1-\alpha/2)$ quantile  of the standard normal distribution.

\end{description}

\section{Numerical studies}\label{sec5}

\subsection{Choice of  smoothing parameters}\label{subsec51}

We develop a data-driven procedure  to
choose the smoothing parameters $L_k$ for $k=1,\ldots,p$, where $L_k$
control the smoothness of $\beta_k(u)$. In practice, various smoothing methods  can be applied to select the smoothing parameters, such as the
cross validation (CV), the generalized cross validation  or the Bayesian information criterion.  Following  \cite{Huang2002}, we propose  a modified ``leave-one-subject-out" CV
to automatically select the smoothing parameters $L_k$ by minimizing the following  CV score:
\begin{equation}\label{CV}
{\rm CV}=\sum_{i=1}^{N}({\bm Y}_i-{\bm
R}_i\hat{\bm\gamma}^{(-i)})^{\tau}M_{\hat{\bm F}^{(-i)}}({\bm Y}_i-{\bm
R}_i\hat{\bm\gamma}^{(-i)}),
\end{equation}
where
$\hat{\bm\gamma}^{(-i)}$ and $\hat{\bm F}^{(-i)}$ are the
estimators defined by solving the nonlinear equations (\ref{equation-1}) and (\ref{equation-2})  from data with the $i$th subject deleted. 
In fact, the CV score  in \eqref{CV} can also be viewed as
a weighted estimate of the true prediction error. 
The performance of the
modified ``leave-one-subject-out" CV procedure  will be evaluated in the next section.

\subsection{Simulation studies}\label{subsec52}

  In this section, we conduct  simulation
studies to assess the finite sample performance of our proposed
methods. The data are  generated from the following model:
\begin{eqnarray}\label{Sim1}
Y_{it}=\beta_1(U_{it})+X_{it}\beta_2(U_{it})+\lambda_i^{\tau}F_t+\varepsilon_{it},
\quad i=1,\ldots,N,\quad t=1,\ldots,T,
\end{eqnarray}
where $\lambda_i=(\lambda_{i1},\lambda_{i2})^{\tau}$,
$F_t=(F_{t1},F_{t2})^{\tau}$, $\beta_1(u)=2-5u+5u^2,
\beta_2(u)=\sin(u\pi),$ $ U_{it}=\omega_{it}+\omega_{i,t-1}$, and
$\omega_{it}$ are i.i.d. random errors from the uniform distribution on  $[0,1/2]$. As the
regressors $X_{it}$ are correlated with $\lambda_i$, $F_t$ and their product $\lambda_i^{\tau}F_t$,  we generate them
according to
\begin{eqnarray*}\label{regressor}
X_{it}=1+\lambda_i^{\tau}F_t+\iota^{\tau}\lambda_i+\iota^{\tau}F_t+\eta_{it},
\end{eqnarray*}
where $\iota=(1,1)^{\tau}$, the effects $\lambda_{ij}, F_{tj}$, $j=1,2$, and
$\eta_{it}$ are all independently from $N(0,1)$. Lastly,  the regression error
$\varepsilon_{it}$ are generated   i.i.d. from  $N(0,4)$. 

As a standard measure of the estimation accuracy, the performance of the estimator $\hat{\bm\beta}(\cdot)$ will be
assessed by  the  integrated squared error (ISE):
\begin{eqnarray*}\label{ISE}
{\rm
ISE}(\hat{\beta}_k)=\int\{\hat{\beta}_k(u)-\beta_k(u)\}^2f(u){\rm d}u,\quad
k=1,2.
\end{eqnarray*}
We further approximate the ISE by the average mean squared error (AMSE):
\begin{eqnarray}\label{AMSE}
{\rm
AMSE}(\hat{\beta}_k)=\frac{1}{NT}\sum_{i=1}^{N}\sum_{t=1}^{T}[\hat{\beta}_k(U_{it})-\beta_k(U_{it})]^2, \quad k=1,2.
\end{eqnarray}
 Throughout the simulations, we  use the cubic
B-spline as the basis functions. Thus $L_k=l_k+m+1$, where $l_k$ is
the number of interior knots and $m=3$ is the degree of the spline. For simplicity, we use  the equally spaced knots for all
numerical studies. To implement the estimation procedure  in Section \ref{sec2}, we   select $L_{k}$ by minimizing the modified
``leave-one-subject-out" CV score in \eqref{CV}. For comparison, we compute the AMSEs in \eqref{AMSE} by three estimation procedures, and report their  numerical results  in Table \ref{tab1}
based on 1000 repetitions.
The column with label  ``IE" denotes the infeasible estimators, which are obtained by assuming observable $F_{t}$.
The column with label ``IFE" denotes  the interactive fixed effects estimators obtained by  our proposed procedure in Section 2. Finally,
the column with label ``LSDVE" denotes the least squares dummy variable estimators, which are obtained under the false assumption with  additive fixed effects in model  \eqref{Sim1} by  applying the  least squares dummy variable method in Section \ref{sec4}.

\begin{table}[h]
\centering
\caption{\label{tab1}  Finite sample performance of the estimators for  model \eqref{Sim1}. }  \tabcolsep 0.1cm
\begin{tabular}{ cccccccc }
\LL && \multicolumn{2}{c}{IE}&
\multicolumn{2}{c}{IFE} &
\multicolumn{2}{c}{LSDVE}\NN
  \cmidrule(r){3-4}\cmidrule(r){5-6}\cmidrule(r){7-8}
$N$& $T$ &  ${\rm AMSE}(\hat{\beta}_1)$& ${\rm AMSE}(\hat{\beta}_2)$ & ${\rm AMSE}(\hat{\beta}_1)$ & ${\rm AMSE}(\hat{\beta}_2)$
& ${\rm AMSE}(\hat{\beta}_1)$ &${\rm AMSE}(\hat{\beta}_2)$
  \ML
  100& 15&  0.0091  &  0.0092& 0.0102   & 0.0103 &0.0947   & 0.0918\\
  100& 30&  0.0045  &  0.0044&   0.0047 &   0.0048 &  0.0878 &   0.0909\\
  100& 60&  0.0021  &  0.0020& 0.0022   & 0.0022 &0.0844   & 0.0829\\
  100& 100&  0.0012  &  0.0012&  0.0013  &  0.0013  &0.0830  &  0.0822\\
  60& 100&  0.0020  &  0.0020 &  0.0021  &  0.0022 & 0.0848 &   0.0838\\
  30& 100&  0.0043  &  0.0042 & 0.0047  &  0.0048  & 0.0864 &   0.0873\\
  15& 100&  0.0082  &  0.0083 & 0.0102  &  0.0102 & 0.0946  & 0.0910   \LL
\end{tabular}%
\end{table}

From Table \ref{tab1}, we note that both the infeasible estimators
and the interactive fixed effects estimators are consistent, and the results of the latter are
gradually closer to those of the former as both $N$ and $T$ increase. However,
the least squares dummy variable estimators of the coefficient functions are biased and inconsistent. One possible reason is that
 the interactive fixed effects are correlated with the
regressors and cannot be removed by the least squares dummy variable method. In addition, AMSEs decrease significantly as both $N$ and $T$ increase for the infeasible estimators and the interactive fixed effects estimators.

\begin{figure}[h!]
\centering
    \includegraphics[width=14cm,height=5cm]{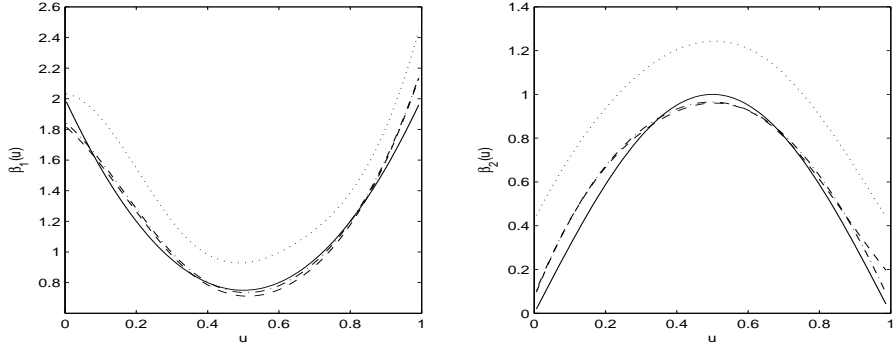}
\caption{Simulation results for model \eqref{Sim1} when $N=100$, $T=60$. In each plot,
the solid curves are for the true coefficient functions,
the dash-dotted curves are for the interactive fixed effects estimators (IFE),
the dashed curves are for the infeasible estimators (IE),
the dotted curves are for the least squares dummy variable estimators (LSDVE).
}
   \label{fig.1}
\end{figure}

Figure \ref{fig.1} presents the  estimated curves  of $\beta_{1}(\cdot)$ and $\beta_{2}(\cdot)$ from a typical
sample, in which the typical sample is selected such that its AMSE is equal to the median of the 1000 replications.
It is also found that the infeasible estimators and the interactive fixed effects estimators are close to the true coefficient functions,
whereas the least squares dummy variable estimators are biased.

\begin{figure}[h!]
    \centering
    \includegraphics[width=14cm,height=5cm]{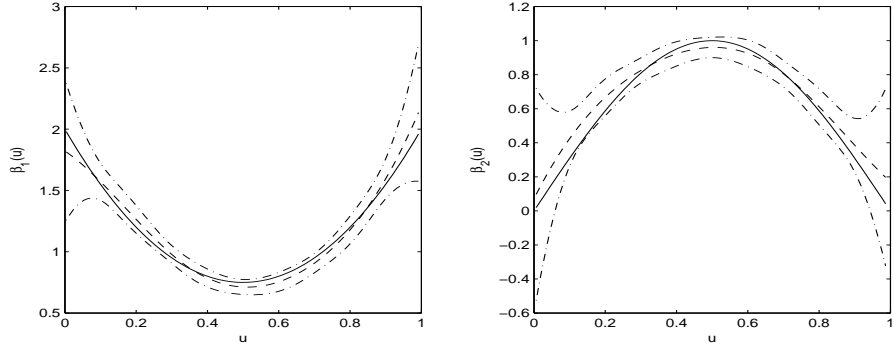}
\caption{ $95\%$ pointwise confidence intervals for ${\bm\beta}(\cdot)$
when $N=100$, $T=60$. In each plot,
the solid curves are for the true coefficient functions,
the dashed curves are for the interactive fixed effects estimators,
the dash-dotted curves are for the $95\%$ pointwise confidence intervals based on bootstrap procedure.
}
   \label{fig.2}
\end{figure}

To construct the $95\%$  pointwise confidence intervals for  $\beta_{1}(\cdot)$ and $\beta_{2}(\cdot)$ using the residual-based block bootstrap procedure in Section \ref{secboot}, we generate 1000 bootstrap samples based on the typical sample, and we choose the block length $l$ by the criterion $l= T^{1/3}$.
The $95\%$ bootstrap pointwise confidence intervals  of $\beta_{1}(\cdot)$ and $\beta_{2}(\cdot)$
are given in Figure \ref{fig.2}. Overall,  the proposed residual-based block bootstrap procedure works quite  well.

\begin{figure}[h!]
    \centering
    \includegraphics[width=14cm,height=5cm]{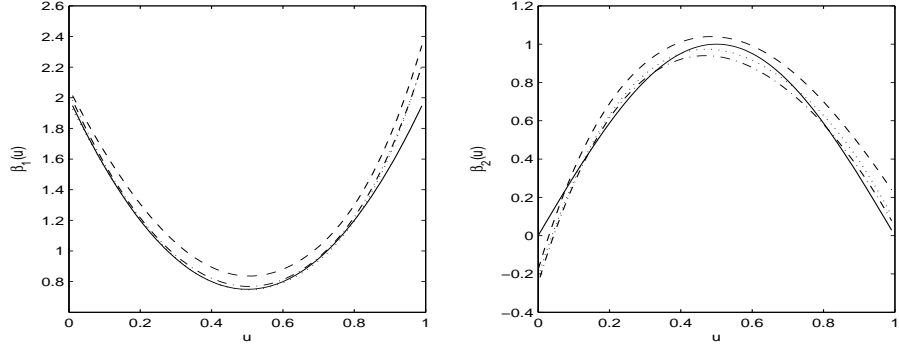}
\caption{ Simulation results for model \eqref{Sim2} with additive fixed effects when $N=100$, $T=60$. In each plot,
the solid curves are for the true coefficient functions,
the dash-dotted curves are for the interactive fixed effects estimators,
the dashed curves are for the infeasible estimators,
the dotted curves are for the least squares dummy variable estimators.
}
   \label{fig.3}
\end{figure}

\begin{figure}[hbtp]
    \centering
    \includegraphics[width=14cm,height=5cm]{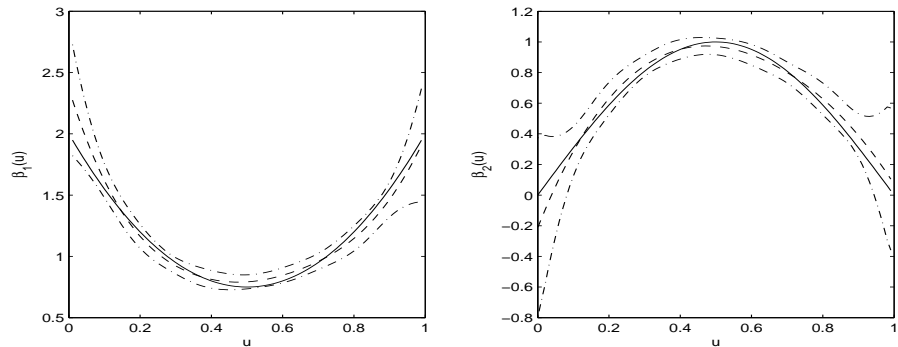}
\caption{$95\%$ pointwise confidence intervals for ${\bm\beta}(\cdot)$
when $N=100$, $T=60$. In each plot,
the solid curves are for the true coefficient functions,
the dashed curves are for the interactive fixed effects estimators,
the dash-dotted curves are for the $95\%$ pointwise confidence intervals based on bootstrap procedure.
}
   \label{fig.4}
\end{figure}

 Our next study is to investigate the performance of  our proposed methods when the fixed effects are additive. Letting $\lambda_{i}=(\mu_{i},1)^{\tau}$ and
$F_{t}=(1,\xi_{t})^{\tau}$,  we have $\lambda_{i}^{\tau}F_{t}=\mu_{i}+\xi_{t}$. We then consider the following varying coefficient panel data model with additive fixed effects:
\begin{eqnarray}\label{Sim2}
Y_{it}=\beta_1(U_{it})+X_{it}\beta_2(U_{it})+\mu_i+\xi_t+\varepsilon_{it},
\quad i=1,\ldots,N,\quad t=1,\ldots,T,
\end{eqnarray}
where $\beta_1(u)$, $\beta_2(u)$,  $U_{it}$ and $\varepsilon_{it}$ are the same as  those in model \eqref{Sim1}. The regressors $X_{it}$ are generated according to
\begin{eqnarray*}\label{X-sim2}
X_{it}=2+2\mu_i+2\xi_t+\eta_{it},
\end{eqnarray*}
where $\eta_{it}\sim N(0,1)$, and the fixed effects are generated by
$$\mu_i\sim N(0,1),\quad i=2,\ldots,N\quad\hbox{and} \quad \mu_1=-\sum_{i=2}^{N}\mu_i,$$
$$\xi_t\sim N(0,1),\quad t=2,\ldots,T\quad\hbox{and} \quad \xi_1=-\sum_{t=2}^{T}\xi_t.$$
With  1000 repetitions, we report the simulation results  in Table \ref{tab2}, Figure \ref{fig.3} and Figure \ref{fig.4}, respectively. To be specific,  Table \ref{tab2} presents the finite sample performance of the estimators for  model \eqref{Sim2} with additive fixed effects, Figure \ref{fig.3} displays the estimated curves of the three estimators for the coefficient functions, and Figure \ref{fig.4} displays the $95\%$ bootstrap pointwise confidence intervals for  ${\beta}_1(\cdot)$ and ${\beta}_2(\cdot)$
when $N=100$ and $T=60$.

\begin{table}[htb!]
 \centering
\caption{\label{tab2}  Finite sample performance of the estimators for  model \eqref{Sim2} with additive fixed effects. } \tabcolsep 0.1cm      
\begin{tabular}{ cccccccc }
\LL && \multicolumn{2}{c}{IE}&
\multicolumn{2}{c}{IFE} &
\multicolumn{2}{c}{LSDVE}\NN
  \cmidrule(r){3-4}\cmidrule(r){5-6}\cmidrule(r){7-8}
$N$& $T$ &  ${\rm AMSE}(\hat{\beta}_1)$& ${\rm AMSE}(\hat{\beta}_2)$ & ${\rm AMSE}(\hat{\beta}_1)$ & ${\rm AMSE}(\hat{\beta}_2)$
& ${\rm AMSE}(\hat{\beta}_1)$ &${\rm AMSE}(\hat{\beta}_2)$
  \ML
  100& 15&  0.0102  &  0.0102& 0.0267  &  0.0260 &0.0083  &  0.0083\\
  100& 30&  0.0048  &  0.0048&   0.0224  &  0.0216 &  0.0040  &  0.0040\\
  100& 60&  0.0022  &  0.0023&0.0192  &  0.0198& 0.0020  &  0.0019\\
  100& 100&  0.0013 &   0.0013&0.0171  &  0.0176&  0.0011  &  0.0011\\
  60& 100&  0.0022  &  0.0022 &  0.0214  &  0.0226 & 0.0019  &  0.0019\\
  30& 100&  0.0046  & 0.0045 & 0.0271   & 0.0281  &  0.0040  &  0.0040\\
  15& 100&  0.0089  &  0.0090 & 0.0340  &  0.0343 & 0.0083  &  0.0083  \LL
\end{tabular}%
\end{table}

Table \ref{tab2} and Figure \ref{fig.3} show that the infeasible estimators, the
interactive fixed effects estimators,  and the least squares dummy variable estimators
are all consistent. Our proposed interactive fixed effects estimators remain
valid even for the varying coefficient panel data model with   additive fixed effects. However, they are less efficient than the least squares dummy variable estimators.
Finally,  the $95\%$ bootstrap pointwise confidence intervals for the typical estimates of $\beta_{1}(\cdot)$ and $\beta_{2}(\cdot)$ in Figure \ref{fig.4} demonstrate the validity
and effectiveness of {our proposed methods}.

\section{Application to a real dataset}\label{sec7}

We apply our proposed methods to a real dataset from the UK  Met Office that contains the monthly mean maximum temperatures (in
Celsius degrees), the mean minimum temperatures (in Celsius degrees), the days of air frost (in days), the total rainfall (in millimeters),
and the total sunshine duration (in hours) from 37 stations. For this dataset, one  main goal  is to investigate the impact of other factors
on the mean maximum temperatures across different stations. \cite{Li2011} analyzed the effect of the total rainfall and the sunshine
duration on the mean maximum temperatures. By contrast, we take into account the days of air frost. Data from 21 stations
during the period of January 2005 to December 2014 are selected while, as the record values for the other stations missed too much, we drop them from further analysis.

Because there exists the seasonal variation in this dataset, our first step is to remove the seasonality from the observations. We impose the
additive decomposition on time series objects and then subtract the seasonal term from the corresponding time series objects. Let
$Y_{it}$ be the seasonally adjusted monthly mean maximum temperatures in the $t$th month in station $i$,
$X_{it,1}$ be the seasonally adjusted monthly days of air frost,
$X_{it,2}$ be the seasonally adjusted monthly total rainfall, and
$X_{it,3}$ be the seasonally adjusted monthly total sunshine duration.
To analyze the dataset, we  consider the following varying coefficient panel data model with interactive fixed effects:
\begin{eqnarray}\label{VCMreal}
Y_{it}=X_{it,1}\beta_{1}(t/T)+X_{it,2}\beta_{2}(t/T)+X_{it,3}\beta_{3}(t/T)+\lambda_i^{\tau}F_t+\varepsilon_{it},
\end{eqnarray}
where $ 1\leq i\leq 21$, $1\leq t\leq 120$, and the multi-factor error structure $\lambda_i^{\tau}F_t+\varepsilon_{it}$ is used to control the heterogeneity
and to capture the unobservable common effects.

\begin{figure}[h!]
    \centering
    \includegraphics[width=12cm,height=9cm]{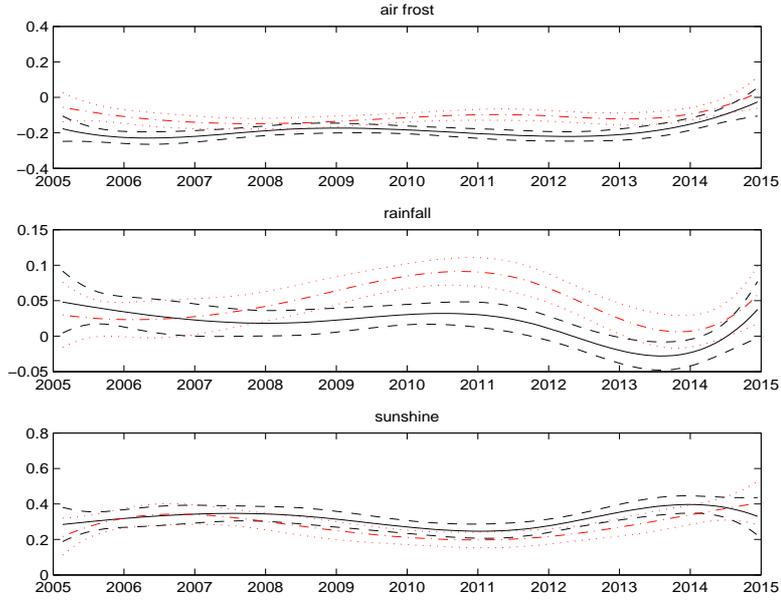}
\caption{The estimated curves and 95\% pointwise confidence intervals of $\beta_{1}(\cdot)$, $\beta_{2}(\cdot)$ and $\beta_{3}(\cdot)$.
In each plot, the solid curves are for the interactive fixed effects estimators,
the dash-dotted curves are for the least squares dummy variable estimators.
The dashed and dotted curves denote the $95\%$ pointwise confidence intervals, respectively.
}
   \label{fig.5}
\end{figure}

Note that the objectives of the study are to estimate the trend effects of the days
of air frost, the monthly total rainfall and the sunshine duration over time.
To achieve the goals, we  fit model (\ref{VCMreal}) using the cubic splines with equally spaced knots,
and select the numbers of interior knots for the unknown coefficient functions
 by minimizing the modified
``leave-one-subject-out" CV score in \eqref{CV}.  To determine the number $r$ of the factors,     we adopt  the  Bayesian information criterion (BIC)  in \cite{Li2016}:
\begin{eqnarray}\label{BIC-fator}
{\rm BIC}(r)=\ln(V(r,\dot{\bm\gamma}_{r}))+r\frac{(N+T)\sum_{k=1}^{p}L_{k}}{NT}\ln\bigg(\frac{NT}{N+T}\bigg),
\end{eqnarray}
where $\dot{\bm\gamma}_{r}$ is the estimator of $\bm\gamma$, and $V(r,\dot{\bm\gamma}_{r})$ is defined as
\begin{eqnarray}\label{V-r}
V(r,\dot{\bm\gamma}_{r})=\frac{1}{NT}\sum_{\varrho=r+1}^{T}\mu_{\varrho}\bigg(\sum_{i=1}^{N}( {\bm Y}_i-{\bm R}_i\dot{\bm\gamma}_{r})({\bm Y}_i-{\bm R}_i\dot{\bm\gamma}_{r})^{\tau}\bigg).
\end{eqnarray}
In \eqref{V-r}, $\mu_{\varrho}(A)$ denotes the $\varrho$-th largest eigenvalue of a symmetric matrix $A$ by counting
multiple eigenvalues multiple times. We set $r_{\max}=8$, and choose  the number $r$ of
the factors by minimizing the objective function  ${\rm BIC}(r)$  in \eqref{BIC-fator}, that is,  $\hat{r}=\arg\min_{0\leq r \leq r_{\max}}{\rm BIC}(r)$.
The estimated curves and $95\%$ bootstrap  pointwise confidence intervals of $\beta_{1}(\cdot)$, $\beta_{2}(\cdot)$ and $\beta_{3}(\cdot)$
are plotted in Figure \ref{fig.5} based on the proposed methods.

From Figure \ref{fig.5},  it is evident  that the estimated curves of
$\beta_{1}(\cdot)$, $\beta_{2}(\cdot)$ and $\beta_{3}(\cdot)$ are all oscillating over time.
Specifically,  the effect of the monthly total sunshine duration is obviously above zero, which shows
 that the monthly total sunshine duration has an overall positive effect on the monthly mean maximum temperatures.
By contrast, we note that the effect of the days of air frost is generally below zero,  which
indicates that there is an overall negative relationship between the monthly mean maximum temperatures
and the days of air frost.

\section{Concluding remarks}\label{sec8}

In this paper, we  use the B-spline
approximations  to study  the  varying coefficient panel data model
with interactive fixed effects.
With an appropriate choice of the smoothing parameters, we propose a robust nonlinear iteration scheme based on the least squares method to estimate the coefficient functions, and then establish the asymptotic theory for
the resulting estimators under some regularity   assumptions,
including the consistency, the convergence rate and the asymptotic distribution.
For the special varying coefficient panel data model with  additive fixed effects, we also develop the least squares dummy variable method to avoid estimating the fixed effects.
In addition, to deal with  the serial correlations within our model that will increase the computational burden and cause the accumulative errors, we propose the
residual-based block bootstrap procedure to construct the pointwise confidence intervals
for the coefficient functions. Simulation studies are also carried out to demonstrate the satisfactory
performance of our proposed methods in practice and to also support the derived
theoretical results.

We note, however, that there still remain two limitations in our paper. First, we assume the cross-sectional independence
that has significantly simplified the theoretical analysis.
Second,  our proposed interactive fixed effects estimators are less efficient than the least squares dummy variable estimators for the varying coefficient panel data model with
 additive fixed effects. In our future  study, we plan to overcome these two limitations by considering  
 the case with cross-sectional dependence, and by testing the   
 additive fixed effects against the interactive fixed effects for the given model.

\section*{Acknowledgements}

Sanying Feng's research was supported by the National Nature Science Foundation of
China (No. 11501522), the Startup Research Fund of Zhengzhou University (No. 1512315004),
Excellent Youth Foundation of Zhengzhou University (No. 32210452).
Gaorong Li's research was supported by the National Natural Science Foundation of China (No. 11471029) and the Beijing Natural Science Foundation (No. 1182003).
 Heng Peng's research was supported by CERG grants from the
Hong Kong Research Grants Council (HKBU 201610 and HKBU 201809), FRG
grants from Hong Kong Baptist University (FRG/08-09/II-33) and the
National Nature Science Foundation of China (10871054). 
Tiejun Tong¡¯s research was supported by the National Natural Science Foundation of China grant (No. 11671338), and the Hong Kong Baptist University grants FRG1/16-17/018 and FRG2/16-17/074.

\appendix
\section*{Appendix A:  Proofs of main results} \label{appA}

\renewcommand{\theequation}{A.\arabic{equation}}
\setcounter{equation}{0}

We provide the proofs of  Theorems \ref{theo1}--\ref{theo5} in  Appendix A. To save space, Lemmas \ref{lemB01}--\ref{lemB6} and their proofs  are provided in Appendix B.


For the ease of the presentation, let $C$ denote some positive constants not
depending on $N$ and $T$, but which may assume different values at
each appearance. In the proof, we use the following properties of
B-spline (see \cite{deBoor2001}): (1) $B_{kl}(u)\geq0$ and
$\sum_{l=1}^{L_k}B_{kl}(u)=1$ for $u\in\mathcal{U}$ and
$k=1,\ldots,p$. (2) There exist constants $0<M_1,M_2<\infty$, not
depending on $L_k$, such that
\begin{eqnarray*}
M_1L_k^{-1}\sum_{l=1}^{L_k}\gamma_{kl}^2\le
\int_{\mathcal{U}}\Big[\sum_{l=1}^{L_k}\gamma_{kl}B_{kl}(u)\Big]^2{\rm d}u\le
M_2L_k^{-1}\sum_{l=1}^{L_k}\gamma_{kl}^2
\end{eqnarray*}
for any sequence $\{\gamma_{kl}\in \mathbb{R}: l=1,\ldots,L_k\}.$

From Assumptions (A1)--(A4) and Corollary 6.21 in \cite{Schumaker1981},
there exists a constant $M$ such that
\begin{eqnarray}\nonumber
\beta_k(u)&=&\sum_{l=1}^{L_k}\widetilde{\gamma}_{kl}B_{kl}(u)+Re_k(u),\\\label{term}
\sup_{u\in\mathcal{U}}|Re_k(u)|&\le& ML_k^{-d},\quad k=1,\ldots,p.
\end{eqnarray}

Let ${\bm e}_i=(e_{i1},\ldots,e_{iT})^{\tau}$ with
$e_{it}=\sum_{k=1}^pRe_k(U_{it})X_{it,k}$ and
$\widetilde{\bm\gamma}=(\widetilde{\bm\gamma}_1^{\tau},\ldots,\widetilde{\bm\gamma}_p^{\tau})^{\tau}$ with
$\widetilde{\bm\gamma}_k=(\widetilde{\gamma}_{k1},\ldots,\widetilde{\gamma}_{kL_{k}})^{\tau}$.
Then ${\bm Y}_i={\bm
R}_i\widetilde{\bm\gamma}+{\bm F}\lambda_i+{\bm \varepsilon}_i+{\bm
e}_i$ for $ i=1,\ldots,N$.  We  use the following facts throughout the paper: $\|\bm F\|=O_P(T^{1/2})$, $\|{\bm R}_i\|=O_P(T^{1/2})$
for all $i$, and $(NT)^{-1}\sum_{i=1}^{N}\|{\bm R}_i\|^2=O_P(1)$. Note that $\|\hat{\bm
F}\|=T^{1/2}\sqrt{r}$. For ease of notation, we define
$\delta_{NT}=\min[\sqrt{N},\sqrt{T}]$ and  $\zeta_{Ld}=\sum_{k=1}^pL_k^{-2d}$. Following the notation of
\cite{Huang2004}, we write $a_n\asymp b_n$ if both $a_n$ and $b_n$ are
positive and $a_n/b_n$ and $b_n/a_n$ are bounded for all $n$.

\vskip12pt

\noindent{\bf Proof.}\ \ We only give the proof of $\|{\bm
R}_i\|=O_P(T^{1/2})$, and omit the proofs of $\|\bm F\|=O_P(T^{1/2})$ and
$(NT)^{-1}\sum_{i=1}^{N}\|{\bm R}_i\|^2=O_P(1)$.
\begin{eqnarray*}
E(\|{\bm R}_i\|^2)&=&
E\Big({\rm tr}({\bm R}_i{\bm R}_i^{\tau})\Big)
=E\Big(\sum_{t=1}^T\|X_{it}^{\tau}{\bm
B}(U_{it})\|^2\Big)\\
&=&E\Big(\sum_{t=1}^T\sum_{k=1}^p\sum_{l=1}^{L_k}X_{it,k}^2B_{kl}^2(U_{it})\Big)\\
&=&\sum_{t=1}^T\sum_{k=1}^p\sum_{l=1}^{L_k}E\Big(X_{it,k}^2B_{kl}^2(U_{it})\Big).
\end{eqnarray*}
By  Assumption (A1), we have
$E\Big(X_{it,k}^2B_{kl}^2(U_{it})\Big)\le C E\Big(B_{kl}^2(U_{it})\Big)$.
Moreover, by the properties of B-spline, we can get that
$$\sum_{l=1}^{L_k}B_{kl}^2(u)\leq \Big(\sum_{l=1}^{L_k}B_{kl}(u)\Big)^2=1.$$
Then we have
$E(\|{\bm R}_i\|^2)=O(T)$, which implies that $\|{\bm
R}_i\|=O_P(T^{1/2})$ for all $i$.

\vskip12pt

\noindent{\bf Proof of Theorem \ref{theo1}.}\ \ Without loss of
generality, assume that $\bm\beta(\cdot)=0$, then
 $\bm {Y}_i={\bm F}\lambda_i+\bm\varepsilon_i$ for $
i=1,\ldots,N$. By Lemma \ref{lemB1} in  Appendix B,  we have
\begin{eqnarray*}
Q_{NT}({\bm\gamma},{\bm F})&=&\frac{1}{NT}\sum_{i=1}^{N}( {\bm
Y}_i-{\bm R}_i\bm\gamma)^{\tau}M_{\bm F}({\bm Y}_i-{\bm
R}_i\bm\gamma)\\
&=&{\bm\gamma}^{\tau}\left(\frac{1}{NT}\sum_{i=1}^{N}{\bm
R}_i^{\tau}M_{\bm F}{\bm R}_i\right){\bm\gamma}+{\rm
tr}\left[\left(\frac{{\bm F}^{\tau}M_{\bm F}{\bm
F}}{T}\right)\left(\frac{\Lambda^{\tau}\Lambda}{N}\right)\right]\\
&&-\frac{2}{NT}{\bm\gamma}^{\tau}\sum_{i=1}^{N}{\bm
R}_i^{\tau}M_{\bm F}{\bm
F}\lambda_i-\frac{2}{NT}{\bm\gamma}^{\tau}\sum_{i=1}^{N}{\bm
R}_i^{\tau}M_{\bm F}\bm\varepsilon_i
\\
&&+\frac{2}{NT}\sum_{i=1}^{N}\lambda_i^{\tau}{\bm F}^{\tau}M_{\bm
F}\bm\varepsilon_i+\frac{1}{NT}\sum_{i=1}^{N}\bm\varepsilon_i^{\tau}M_{\bm
F}\bm\varepsilon_i\\
&=:&\widetilde{Q}_{NT}({\bm\gamma},{\bm F})+o_P(1)
\end{eqnarray*}
uniformly over bounded $\bm\gamma$ and over $\bm F$ such that ${\bm
F}^{\tau}{\bm F}/T=I$, where
\begin{eqnarray*}
\widetilde{Q}_{NT}({\bm\gamma},{\bm
F})&=&{\bm\gamma}^{\tau}\left(\frac{1}{NT}\sum_{i=1}^{N}{\bm
R}_i^{\tau}M_{\bm F}{\bm R}_i\right){\bm\gamma}+{\rm
tr}\left[\left(\frac{{\bm F}^{\tau}M_{\bm F}{\bm
F}}{T}\right)\left(\frac{\Lambda^{\tau}\Lambda}{N}\right)\right]\\
&&-\frac{2}{NT}{\bm\gamma}^{\tau}\sum_{i=1}^{N}{\bm
R}_i^{\tau}M_{\bm F}{\bm F}\lambda_i.
\end{eqnarray*}
Let $\eta={\rm vec}(M_{\bm F}{\bm F}),$ and
\begin{eqnarray*}
A_{1}=\frac{1}{NT}\sum_{i=1}^{N}{\bm R}_i^{\tau}M_{\bm F}{\bm R}_i,\quad
A_{2}=\left(\frac{\Lambda^{\tau}\Lambda}{N}\otimes I_T\right),\quad
A_{3}=\frac{1}{NT}\sum_{i=1}^{N}(\lambda_i^{\tau}\otimes M_{\bm F}{\bm
R}_i).
\end{eqnarray*}
 Then,
\begin{eqnarray*}
\widetilde{Q}_{NT}({\bm\gamma},{\bm
F})&=&{\bm\gamma}^{\tau}A_{1}{\bm\gamma}+\eta^{\tau}A_{2}\eta-2{\bm\gamma}^{\tau}A_{3}^{\tau}\eta\\
&=&{\bm\gamma}^{\tau}(A_{1}-A_{3}^{\tau}A_{2}^{-1}A_{3}){\bm\gamma}+(\eta^{\tau}-{\bm\gamma}^{\tau}A_{3}^{\tau}A_{2}^{-1})A_{2}(\eta-A_{2}^{-1}A_{3}{\bm\gamma})\\
&=:&{\bm\gamma}^{\tau}D(\bm F){\bm\gamma}+\theta^{\tau}A_{2}\theta,
\end{eqnarray*}
where $\theta=\eta-A_{2}^{-1}A_{3}{\bm\gamma}.$ By Assumption (A5), $D(\bm
F)$ is a positive definite matrix and $A_{2}$ is also a positive
definite matrix, which show that $\widetilde{Q}_{NT}({\bm\gamma},{\bm F})\geq0.$
By the similar argument as in \cite{Bai2009}, it is easy to show that
$\widetilde{Q}_{NT}({\bm\gamma},{\bm F})$ achieves its unique
minimum at $(0,{\bm F}H)$ for any $r\times r$ invertible matrix
$H$.  Thus, $\hat{\beta}_k(\cdot), k=1,\ldots,p,$ are uniquely defined. This completes the proof of part (i).

The proof of (ii) is similar to that of Proposition 1 (ii) in \cite{Bai2009}. To save space, we do not present the detailed proof.
\hfill$\Box$

\vskip12pt
\noindent{\bf Proof of Theorem \ref{theo2}.}\ \ Since
$\hat{\beta}_k(u)=\sum\limits_{l=1}^{L_k}\hat{\gamma}_{kl}{B}_{kl}(u)$ and
$\widetilde{\beta}_k(u)=\sum\limits_{l=1}^{L_k}\widetilde{\gamma}_{kl}{B}_{kl}(u)$,
 by the
properties of B-spline and  (\ref{term}), we have
\begin{eqnarray*}\label{Inq1}
\|\hat{\beta}_k(\cdot)-{\beta}_k(\cdot)\|_{L_2}^2 \le
2\|\hat{\beta}_k(\cdot)-\widetilde{\beta}_k(\cdot)\|_{L_2}^2+ML_k^{-2d}
\end{eqnarray*}
and
\begin{eqnarray}\label{Inq2}
\|\hat{\beta}_k(\cdot)-\widetilde{\beta}_k(\cdot)\|_{L_2}^2=\|\hat{\bm\gamma}_k-\widetilde{\bm\gamma}_k\|_H^2\asymp
L_k^{-1}\|\hat{\bm\gamma}_k-\widetilde{\bm\gamma}_k\|^2,\quad
k=1,\ldots,p,
\end{eqnarray}
where $\|\bm\gamma_k\|_H^2=\bm\gamma_k^{\tau}{\bm
H}_{k}\bm\gamma_k$, and ${\bm H}_k=(h_{ij})_{L_{k}\times L_{k}}$ is
a matrix with entries
$h_{ij}=\int_{\mathcal{U}}B_{ki}(u)B_{kj}(u){\rm d}u$. Summing over $k$ for
(\ref{Inq2}), we obtain that
\begin{eqnarray*}
\|\hat{\bm\beta}(\cdot)-\widetilde{\bm\beta}(\cdot)\|_{L_2}^2=\sum_{k=1}^p\|\hat{\bm\gamma}_k-\widetilde{\bm\gamma}_k\|_H^2\asymp
L_N^{-1}\|\hat{\bm\gamma}-\widetilde{\bm\gamma}\|^2.
\end{eqnarray*}
By  (\ref{equation-1}) and ${\bm Y}_i={\bm
R}_i\widetilde{\bm\gamma}+{\bm F}\lambda_i+{\bm \varepsilon}_i+{\bm
e}_i$ for $ i=1,\ldots,N$, we have
\begin{eqnarray*}
\hat{\bm\gamma}-\widetilde{\bm\gamma}=\Big(\sum_{i=1}^N{\bm
R}_i^{\tau}M_{\hat{\bm F}}{\bm R}_i\Big)^{-1}\sum_{i=1}^N{\bm
R}_i^{\tau}M_{\hat{\bm F}}({\bm F}\lambda_i+{\bm \varepsilon}_i+{\bm
e}_i),
\end{eqnarray*}
or equivalently,
\begin{eqnarray}\label{A-1}
\Big(\sum_{i=1}^N{\bm R}_i^{\tau}M_{\hat{\bm F}}{\bm
R}_i\Big)(\hat{\bm\gamma}-\widetilde{\bm\gamma}) =\sum_{i=1}^N{\bm
R}_i^{\tau}M_{\hat{\bm F}}{\bm F}\lambda_i+\sum_{i=1}^N{\bm
R}_i^{\tau}M_{\hat{\bm F}}{\bm\varepsilon}_i+\sum_{i=1}^N{\bm
R}_i^{\tau}M_{\hat{\bm F}}{\bm e}_i.
\end{eqnarray}
We first deal with the third term of the right hand in \eqref{A-1}. By Assumption
(A1) and (\ref{term}), and using the similar proofs to Lemma
A.7 in \cite{Huang2004}, and Lemmas \ref{lemB1} and
\ref{lemB2} in  Appendix B, it is easy to show that
\begin{eqnarray}\label{A-3}
\left\|\frac{1}{NT}\sum_{i=1}^N{\bm R}_i^{\tau}M_{\hat{\bm F}}{\bm
e}_i\right\|^2=O_P\Big(L_N^{-1}\zeta_{Ld}\Big).
\end{eqnarray}
For the first term of the right hand in \eqref{A-1}, by noting that
$M_{\hat{\bm F}}\hat{{\bm F}}=0$, we have $M_{\hat{\bm F}}{\bm
F}=M_{\hat{\bm F}}({\bm F}-\hat{{\bm F}}H^{-1})$. By (\ref{A-2-1})
in  Appendix B, we have
\begin{eqnarray}\label{A-2}
{\bm F}-\hat{\bm F}H^{-1}=-(B_1+B_2+\cdots+B_{15})G,
\end{eqnarray}
where $H=(\Lambda^{\tau}\Lambda/N)({\bm F}^{\tau}\hat{\bm
F}/T)V_{NT}^{-1}$,  $G=({\bm F}^{\tau}\hat{\bm
F}/T)^{-1}(\Lambda^{\tau}\Lambda/N)^{-1}$ is a matrix of fixed
dimension and does not vary with $i$, and $B_1,\ldots,B_{15}$ are
defined in (\ref{A-2-1}) of  Appendix B. By (\ref{A-2}), we have
\begin{eqnarray*}
\frac{1}{NT}\sum_{i=1}^N{\bm R}_i^{\tau}M_{\hat{\bm F}}{\bm
F}\lambda_i&=&\frac{1}{NT}\sum_{i=1}^N{\bm R}_i^{\tau}M_{\hat{\bm
F}}({\bm F}-\hat{{\bm F}}H^{-1})\lambda_i\\
&=&-\frac{1}{NT}\sum_{i=1}^N{\bm R}_i^{\tau}M_{\hat{\bm
F}}(B_1+B_2+\cdots+B_{15})G\lambda_i\\
&=:&J_1+J_2+\cdots+J_{15}.
\end{eqnarray*}
It is easy to see that $J_1$---$J_{15}$ depend on
$B_1$---$B_{15}$ respectively. For $J_2$, we have
\begin{eqnarray*}
J_2&=&-\frac{1}{NT}\sum_{i=1}^N{\bm R}_i^{\tau}M_{\hat{\bm
F}}\left[\frac{1}{NT}\sum_{j=1}^{N}{\bm
R}_j(\widetilde{\bm\gamma}-\hat{\bm\gamma})\lambda_j^{\tau}{\bm
F}^{\tau}\hat{\bm F}\right]\left(\frac{{\bm F}^{\tau}\hat{\bm
F}}{T}\right)^{-1}\left(\frac{\Lambda^{\tau}\Lambda}{N}\right)^{-1}\lambda_i\\
&=&\frac{1}{N^2T}\sum_{i=1}^N\sum_{j=1}^N({\bm
R}_i^{\tau}M_{\hat{\bm F}}{\bm R}_j)
\left[\lambda_j^{\tau}\left(\frac{\Lambda^{\tau}\Lambda}{N}\right)^{-1}\lambda_i\right](\hat{\bm\gamma}-\widetilde{\bm\gamma})\\
&=&\frac{1}{T}\left[\frac{1}{N^2}\sum_{i=1}^N\sum_{j=1}^N{\bm
R}_i^{\tau}M_{\hat{\bm F}}{\bm
R}_ja_{ij}\right](\hat{\bm\gamma}-\widetilde{\bm\gamma}),
\end{eqnarray*}
where
$a_{ij}=\lambda_i^{\tau}(\Lambda^{\tau}\Lambda/N)^{-1}\lambda_j.$
For $J_1$, we have
\begin{eqnarray*}
J_1=-\frac{1}{NT}\sum_{i=1}^N{\bm R}_i^{\tau}M_{\hat{\bm
F}}(B_1)G\lambda_i=o_P(\|\hat{\bm\gamma}-\widetilde{\bm\gamma}\|).
\end{eqnarray*}

For $J_3$, we have
\begin{eqnarray*}
J_3=\frac{1}{N^2T}\sum_{i=1}^N\sum_{j=1}^N{\bm
R}_i^{\tau}M_{\hat{\bm F}}{\bm
R}_j\Big(\frac{\bm\varepsilon_j^{\tau}\hat{\bm
F}}{T}\Big)G\lambda_i(\hat{\bm\gamma}-\widetilde{\bm\gamma}).
\end{eqnarray*}
By Lemma \ref{lemB2} in  Appendix B and some elementary
calculations, we have
\begin{eqnarray}\label{ApendA-1}
T^{-1}\bm\varepsilon_j^{\tau}\hat{\bm
F}&=&T^{-1}\bm\varepsilon_j^{\tau}{\bm
F}H+T^{-1}\bm\varepsilon_j^{\tau}(\hat{\bm F}-{\bm F}H)\\\nonumber
&=&O_P(T^{-1/2})+T^{-1/2}O_P(\|\hat{\bm\gamma}-\widetilde{\bm\gamma}\|)+O_P(\delta_{NT}^{-2})
+O_P\left(\zeta_{Ld}^{1/2}T^{-1/2}\right).
\end{eqnarray}
Using the above result and the similar argument as the proof of Lemma
\ref{lemB1} in  Appendix B, it is easy to verify  that
$J_3=o_P(\|\hat{\bm\gamma}-\widetilde{\bm\gamma}\|).$ Similarly, we
can obtain that
$J_5=o_P(\|\hat{\bm\gamma}-\widetilde{\bm\gamma}\|).$ For $J_4$, we
have
\begin{eqnarray*}
J_4=-\frac{1}{N^2T}\sum_{i=1}^N\sum_{j=1}^N{\bm
R}_i^{\tau}M_{\hat{\bm F}}{\bm
F}\lambda_j(\widetilde{\bm\gamma}-\hat{\bm\gamma})^{\tau}
\Big(\frac{{\bm R}_j^{\tau}\hat{\bm F}}{T}\Big)G\lambda_i.
\end{eqnarray*}
Noting that $M_{\hat{\bm F}}{\bm F}=M_{\hat{\bm F}}({\bm F}-\hat{\bm
F}H^{-1})$, and using Lemma \ref{lemB2} (i) in  Appendix B, that
is, $T^{-1/2}\|{\bm F}-\hat{\bm
F}H^{-1}\|=O_P(\|\hat{\bm\gamma}-\widetilde{\bm\gamma}\|)+O_P(\delta_{NT}^{-1})+O_P(\zeta_{Ld}^{1/2})$,
we can obtain that
$J_4=o_P(\|\hat{\bm\gamma}-\widetilde{\bm\gamma}\|).$ For $J_6$,
noting that $G$ is a matrix of fixed dimension and does not vary
with $i$, and by $M_{\hat{\bm F}}{\bm F}=M_{\hat{\bm F}}({\bm
F}-\hat{\bm F}H^{-1})$, we have
\begin{eqnarray*}
J_6&=&-\frac{1}{N^2T}\sum_{i=1}^N\sum_{j=1}^N{\bm
R}_i^{\tau}M_{\hat{\bm F}}{\bm F}\lambda_j \Big(\frac{{\bm
\varepsilon}_j^{\tau}\hat{\bm F}}{T}\Big)G\lambda_i\\
&=&-\frac{1}{NT}\sum_{i=1}^N{\bm R}_i^{\tau}M_{\hat{\bm F}}({\bm
F}-\hat{\bm F}H^{-1})\Big(\frac{1}{N}\sum_{j=1}^N\lambda_j
\Big(\frac{{\bm \varepsilon}_j^{\tau}\hat{\bm
F}}{T}\Big)\Big)G\lambda_i.
\end{eqnarray*}
By (\ref{ApendA-1}) and Lemma \ref{lemB2} in  Appendix B, we have
\begin{eqnarray*}
\frac{1}{NT}\sum_{j=1}^N\lambda_j {\bm \varepsilon}_j^{\tau}\hat{\bm
F}&=&\frac{1}{NT}\sum_{j=1}^N\lambda_j {\bm
\varepsilon}_j^{\tau}{\bm F}H+\frac{1}{NT}\sum_{j=1}^N\lambda_j {\bm
\varepsilon}_j^{\tau}(\hat{\bm F}-{\bm F}H)\\
&=&O_P((NT)^{-1/2})+(TN)^{-1/2}O_P(\|\hat{\bm\gamma}-\widetilde{\bm\gamma}\|)
+
O_P(N^{-1})\\
&&+N^{-1/2}O_P(\delta_{NT}^{-2})+N^{-1/2}O_P\Big(\zeta_{Ld}^{1/2}\Big)\\
&=&O_P((NT)^{-1/2}) +
O_P(N^{-1})+N^{-1/2}O_P(\delta_{NT}^{-2})\\
&&+N^{-1/2}O_P\Big(\zeta_{Ld}^{1/2}\Big).
\end{eqnarray*}
By  Lemma \ref{lemB2} (v) in  Appendix B, then
$$\dfrac{1}{NT}\sum\limits_{i=1}^N{\bm R}_i^{\tau}M_{\hat{\bm F}}(\hat{\bm
F}-{\bm
F}H)=O_P(\|\hat{\bm\gamma}-\widetilde{\bm\gamma}\|)
+O_P(\delta_{NT}^{-2})+O_P(\zeta_{Ld}^{1/2}).$$ Moreover, the matrix $G$ does not depend on $i$
and $\|G\|=O_P(1)$, then
\begin{eqnarray*}
J_6&=&\Big[O_P(\|\hat{\bm\gamma}-\widetilde{\bm\gamma}\|)
+O_P(\delta_{NT}^{-2})+O_P\Big(\zeta_{Ld}^{1/2}\Big)\Big]\\
&&\times\Big[O_P((NT)^{-1/2}) +
O_P(N^{-1})+N^{-1/2}O_P(\delta_{NT}^{-2})+N^{-1/2}O_P\Big(\zeta_{Ld}^{1/2}\Big)\Big]\\
&=&o_P(\|\hat{\bm\gamma}-\widetilde{\bm\gamma}\|)+o_P((NT)^{-1/2})
+N^{-1}O_P(\delta_{NT}^{-2})+N^{-1/2}O_P(\delta_{NT}^{-4})\\
&&+N^{-1}O_P\Big(\zeta_{Ld}^{1/2}\Big)+
N^{-1/2}O_P\left(\zeta_{Ld}\right).
\end{eqnarray*}

For $J_7$, we have
\begin{eqnarray*}
J_7&=&-\frac{1}{N^2T}\sum_{i=1}^N{\bm R}_i^{\tau}M_{\hat{\bm
F}}\left[\sum_{j=1}^N{\bm\varepsilon}_j\lambda_j^{\tau}\Big(\frac{\Lambda^{\tau}\Lambda}{N}\Big)^{-1}\right]\lambda_i\\
&=&-\frac{1}{N^2T}\sum_{i=1}^N\sum_{j=1}^Na_{ij}{\bm
R}_i^{\tau}M_{\hat{\bm F}}{\bm\varepsilon}_j,
\end{eqnarray*}
where
$a_{ij}=\lambda_i^{\tau}(\Lambda^{\tau}\Lambda/N)^{-1}\lambda_j.$
For $J_8$, by Assumption (A8), and the same argument as in the
Proposition A.2 of \cite{Bai2009}, and Lemma \ref{lemB4} in Appendix
B, 
we have
\begin{small}
\begin{eqnarray*}
J_8&=&-\frac{1}{N^2T^2}\sum_{i=1}^N\sum_{j=1}^N{\bm
R}_i^{\tau}M_{\hat{\bm F}}{\bm \varepsilon}_j {\bm
\varepsilon}_j^{\tau}\hat{\bm F}G\lambda_i\\
&=&-\frac{1}{N^2T^2}\sum_{i=1}^N\sum_{j=1}^N{\bm
R}_i^{\tau}M_{\hat{\bm F}}\Omega_j\hat{\bm
F}G\lambda_i-\frac{1}{N^2T^2}\sum_{i=1}^N\sum_{j=1}^N{\bm
R}_i^{\tau}M_{\hat{\bm F}}({\bm \varepsilon}_j {\bm
\varepsilon}_j^{\tau}-\Omega_j)\hat{\bm F}G\lambda_i\\
&=:&A_{NT}+O_P(1/(T\sqrt{N}))+(NT)^{-1/2}\left[O_P(\|\hat{\bm\gamma}-\widetilde{\bm\gamma}\|)+O_P(\delta_{NT}^{-1})+O_P\Big(\zeta_{Ld}^{1/2}\Big)\right]\\
&&+\frac{1}{\sqrt{N}}\left[O_P(\|\hat{\bm\gamma}-\widetilde{\bm\gamma}\|)+O_P(\delta_{NT}^{-1})+O_P\Big(\zeta_{Ld}^{1/2}\Big)\right]^2.
\end{eqnarray*}
\end{small}
For $J_9$  and $J_{10}$, which depend on
$\hat{\bm\gamma}-\widetilde{\bm\gamma}$. Using the same argument, it
is easy to prove that $J_9$  and $J_{10}$ are bounded in the
Euclidean norm by $o_P(\|\hat{\bm\gamma}-\widetilde{\bm\gamma}\|)$.
For $J_{11}$, using $M_{\hat{\bm F}}{\bm F}=M_{\hat{\bm F}}({\bm
F}-\hat{\bm F}H^{-1})$ again, and letting $\widetilde{\bm W}_j={\bm
e}_j^{\tau}\hat{\bm F}/T$ and $\|\widetilde{\bm W}_j\|=\|{\bm
e}_j\|\sqrt{r}/\sqrt{T}=O_P(\zeta_{Ld}^{1/2})$,
and using Lemma \ref{lemB2} (v) in  Appendix B, we have
\begin{eqnarray*}
J_{11}&=&-\frac{1}{N^2T}\sum_{i=1}^N\sum_{j=1}^N{\bm
R}_i^{\tau}M_{\hat{\bm F}}{\bm F}\lambda_j \Big(\frac{{\bm
e}_j^{\tau}\hat{\bm F}}{T}\Big)G\lambda_i\\
&=&-\frac{1}{NT}\sum_{i=1}^N{\bm R}_i^{\tau}M_{\hat{\bm F}}({\bm
F}-\hat{\bm F}H^{-1})\Big(\frac{1}{N}\sum_{j=1}^N\lambda_j
\Big(\frac{{\bm e}_j^{\tau}\hat{\bm F}}{T}\Big)\Big)G\lambda_i\\
&=&O_P\Big(\zeta_{Ld}^{1/2}\Big)\left[O_P(\|\hat{\bm\gamma}-\widetilde{\bm\gamma}\|)
+O_P(\delta_{NT}^{-2})+O_P\Big(\zeta_{Ld}^{1/2}\Big)\right].
\end{eqnarray*}
For $J_{12}$, similar to (\ref{A-3}), we have
\begin{eqnarray*}
J_{12}&=&-\frac{1}{N^2T}\sum_{i=1}^N{\bm R}_i^{\tau}M_{\hat{\bm
F}}\left[\sum_{j=1}^N{\bm e}_j\lambda_j^{\tau}\Big(\frac{\Lambda^{\tau}\Lambda}{N}\Big)^{-1}\right]\lambda_i\\
&=&-\frac{1}{N^2T}\sum_{i=1}^N\sum_{j=1}^Na_{ij}{\bm
R}_i^{\tau}M_{\hat{\bm F}}{\bm e}_j\\
&=&O_P\Big(L_N^{-1/2}\zeta_{Ld}^{1/2}\Big),
\end{eqnarray*}
where
$a_{ij}=\lambda_i^{\tau}(\Lambda^{\tau}\Lambda/N)^{-1}\lambda_j.$
Using the similar argument, it is easy to see that
$J_{13}=(NT)^{-1/2}O_P(\zeta_{Ld}^{1/2})$.

For $J_{14}$, by (\ref{ApendA-1}) we have
\begin{eqnarray*}
J_{14}&=&-\frac{1}{N^2T}\sum_{i=1}^N\sum_{j=1}^N{\bm
R}_i^{\tau}M_{\hat{\bm F}}{\bm e}_j \Big(\frac{{\bm
\varepsilon}_j^{\tau}\hat{\bm F}}{T}\Big)G\lambda_i\\
&=&-\frac{1}{N^2T}\sum_{i=1}^N\sum_{j=1}^N{\bm
R}_i^{\tau}M_{\hat{\bm F}}{\bm e}_j \Big(\frac{{\bm
\varepsilon}_j^{\tau}{\bm
F}H}{T}\Big)G\lambda_i\\
&&-\frac{1}{N^2T}\sum_{i=1}^N\sum_{j=1}^N{\bm R}_i^{\tau}M_{\hat{\bm
F}}{\bm e}_j \Big(\frac{{\bm \varepsilon}_j^{\tau}(\hat{\bm F}-{\bm
F}H)}{T}\Big)G\lambda_i.
\end{eqnarray*}
Similarly, we can  prove that the first term of the above
equation is bounded by
$T^{-1/2}O_P(\zeta_{Ld}^{1/2})$. For
the second term, by a similar argument and Lemma \ref{lemB3} in
 Appendix B, we can prove that the second term is bounded above
by
$$O_P\Big(\zeta_{Ld}^{1/2}\Big)\Big[T^{-1/2}O_P(\|\hat{\bm\gamma}-\widetilde{\bm\gamma}\|)+O_P(\delta_{NT}^{-2})
+O_P\Big(\zeta_{Ld}^{1/2}T^{-1/2}\Big)\Big].$$


For $J_{15}$, by $M_{\hat{\bm
F}}\hat{\bm F}=0$ and some simple calculations, we have
\begin{eqnarray*}
J_{15}=-\frac{1}{N^2T}\sum_{i=1}^N\sum_{j=1}^N{\bm
R}_i^{\tau}M_{\hat{\bm F}} \Big(\frac{{\bm e}_j{\bm
e}_j^{\tau}}{T}\Big)\hat{\bm F}G\lambda_i
=o_P(\zeta_{Ld}).
\end{eqnarray*}
Summarizing the above results, we can obtain that
\begin{eqnarray*}
\frac{1}{NT}\sum_{i=1}^N{\bm R}_i^{\tau}M_{\hat{\bm F}}{\bm
F}\lambda_i&=&J_2+J_7+A_{NT}+o_P(\|\hat{\bm\gamma}-\widetilde{\bm\gamma}\|)+o_P((NT)^{-1/2})+O_P\Big(\frac{1}{T\sqrt{N}}\Big)
\\&&
 + N^{-1/2}O_P(\delta_{NT}^{-2})+O_P\Big(T^{-1/2}\zeta_{Ld}^{1/2}\Big)+O_P\Big(L_{N}^{-1/2}\zeta_{Ld}^{1/2}\Big).
\end{eqnarray*}
This leads to
\begin{eqnarray*}
&&\left(\frac{1}{NT}\sum_{i=1}^N{\bm R}_i^{\tau}M_{\hat{\bm F}}{\bm
R}_i+o_P(1)\right)(\hat{\bm\gamma}-\widetilde{\bm\gamma})-J_2\\
&=&\frac{1}{NT}\sum_{i=1}^N{\bm R}_i^{\tau}M_{\hat{\bm
F}}{\bm\varepsilon}_i+J_7+A_{NT}+o_P((NT)^{-1/2})
+O_P\Big(\frac{1}{T\sqrt{N}}\Big)\\
&&~~~~~+N^{-1/2}O_P(\delta_{NT}^{-2})+O_P\Big(T^{-1/2}\zeta_{Ld}^{1/2}\Big)+O_P\Big(L_{N}^{-1/2}\zeta_{Ld}^{1/2}\Big).
\end{eqnarray*}
Multiplying $L_N(L_ND(\hat{\bm F}))^{-1}$ on each side of
the above equation, and by Lemma \ref{lemB5} in Appendix B,
\begin{eqnarray*}
\hat{\bm\gamma}-\widetilde{\bm\gamma} &=&\left(L_ND(\hat{\bm
F})\right)^{-1}\frac{L_N}{NT}\sum_{i=1}^N\left[{\bm
R}_i^{\tau}M_{{\bm F}}-\frac{1}{N}\sum_{j=1}^Na_{ij}{\bm
R}_j^{\tau}M_{{\bm
F}}\right]{\bm\varepsilon}_i +\frac{L_N}{T}\Lambda_{NT}\\
&&+\frac{L_N}{N}\left(L_ND(\hat{\bm
F})\right)^{-1}\xi_{NT}^*+\left(L_ND(\hat{\bm
F})\right)^{-1}o_P\left(L_N(NT)^{-1/2}\right)\\
&&+\left(L_ND(\hat{\bm
F})\right)^{-1}O_P\Big(\frac{L_N}{T\sqrt{N}}\Big)
 +L_NN^{-1/2}\left(L_ND(\hat{\bm
F})\right)^{-1}O_P(\delta_{NT}^{-2})\\
&&+\left(L_ND(\hat{\bm
F})\right)^{-1}O_P\Big(L_N T^{-1/2}\zeta_{Ld}^{1/2}\Big)
+\left(L_ND(\hat{\bm
F})\right)^{-1}O_P\Big(L_N^{1/2}\zeta_{Ld}^{1/2}\Big),
\end{eqnarray*}
where
\begin{eqnarray*}
\xi_{NT}^*=-\frac{1}{N}\sum_{i=1}^N\sum_{j=1}^N\frac{({\bm R}_i-{\bm
V}_i)^{\tau}{\bm F}}{T}\left(\frac{{\bm F}^{\tau}{\bm
F}}{T}\right)^{-1}\left(\frac{\Lambda^{\tau}\Lambda}{N}\right)^{-1}\lambda_j
\left(\frac{1}{T}\sum_{t=1}^T\varepsilon_{it}\varepsilon_{jt}\right)=O_P(1)
\end{eqnarray*}
and
\begin{eqnarray*}
\Lambda_{NT}=-\left(L_ND(\hat{\bm
F})\right)^{-1}\frac{1}{NT}\sum_{i=1}^N{\bm R}_i^{\tau}M_{\hat{\bm
F}}\Omega\hat{\bm F}G\lambda_i
\end{eqnarray*}
with $\Omega=\frac{1}{N}\sum_{j=1}^N\Omega_j$ and
$\Omega_j=E(\bm\varepsilon_j\bm\varepsilon_j^{\tau})$.
By Lemmas \ref{lemB01} and \ref{lemB6} in Appendix B, it can be shown that
$D(\hat{\bm F})=D({\bm F})+o_P(1)$ and the minimum and maximum
eigenvalues of $L_ND(\hat{\bm F})$ are bounded with probability
tending to 1. In addition, by Lemma \ref{lemB01} in Appendix B and Lemma A.6 in
\cite{Bai2009}, it is easy to verify  that $\Lambda_{NT}=O_P(1)$. Using
the same argument for Lemma \ref{lemB1}, we have
\begin{eqnarray*}
&&\left\|D(\bm F)^{-1}\frac{1}{NT}\sum_{i=1}^N\left[{\bm
R}_i^{\tau}M_{{\bm F}}-\frac{1}{N}\sum_{j=1}^Na_{ij}{\bm
R}_j^{\tau}M_{{\bm F}}\right]{\bm\varepsilon}_i\right\|^2\\
&\asymp& \left\|\frac{L_N}{NT}\sum_{i=1}^N\left[{\bm
R}_i^{\tau}M_{{\bm F}}-\frac{1}{N}\sum_{j=1}^Na_{ij}{\bm
R}_j^{\tau}M_{{\bm F}}\right]{\bm\varepsilon}_i\right\|^2\\
&=&O_P(L_N^2(NT)^{-1})
\end{eqnarray*}
uniformly for $\bm F$. By the above results, together with   Lemma
\ref{lemB01} and $\delta_{NT}^{-2}L_{N}\log L_{N}\rightarrow0$ as
$N,T\rightarrow\infty$, we have
\begin{eqnarray*}
\|\hat{\bm\gamma}-\widetilde{\bm\gamma}\|&=&O_P(L_N(NT)^{-1/2})+O_P(L_NT^{-1})+O_P(L_NN^{-1})\\
&&+O_P\Big(L_{N}T^{-1/2}\zeta_{Ld}^{1/2}\Big)+O_P\Big(L_N^{1/2}\zeta_{Ld}^{1/2}\Big).
\end{eqnarray*}
\hfill$\Box$

\vskip12pt

\noindent{\bf Proof of Theorem \ref{theo3}.}
Let $\check{\bm\gamma}=E(\hat{\bm\gamma}|\mathcal{D}).$
Using the similar proof to Theorem 4.1 in \cite{Huang2003}, and invoking
Lemma A.8 in \cite{Huang2004} and the proof of Theorem 2
in \cite{Wang2008},
we obtain that, for any vector $c_{n}$ with dimension $\sum_{k=1}^{p}L_{k}$ and whose components are not all zero,
$$\{c_{n}^{\tau}\Phi c_{n}\}^{-1/2}c_{n}^{\tau}(\hat{\bm\gamma}-\check{\bm\gamma})\stackrel{L}\longrightarrow N(0,1).$$
For any $p$-vector $a_n$ whose components are not all zero, letting $c_{n} = {\bm B}(u)^{\tau}a_{n}$, we have
$$\{a_{n}^{\tau}{\rm Var}(\hat{\bm\beta}(u)|\mathcal{D})a_{n}\}^{-1/2}a_{n}^{\tau}(\hat{\bm\beta}(u)-\check{\bm\beta}(u))
\stackrel{L}\longrightarrow N(0,1),$$
which in turn yields the desired result.
\hfill$\Box$

\vskip12pt
\noindent{\bf Proof of Theorem \ref{theo4}.}
Note that
$$\check{\bm\beta}(u)-\bm\beta(u)={\bm B}(u)^{\tau}(\check{\gamma}-\tilde{\gamma})+{\bm B}(u)^{\tau}\tilde{\gamma}-\bm\beta(u).$$
By (\ref{term}), we have $\|{\bm B}(u)^{\tau}\tilde{\gamma}-\bm\beta(u)\|_{\infty}=O_P(\zeta_{Ld}^{1/2})$.
Furthermore, a simple calculation yields
\begin{eqnarray*}
\check{\gamma}-\tilde{\gamma}=\Big(\sum_{i=1}^N{\bm
R}_i^{\tau}M_{\hat{\bm F}}{\bm R}_i\Big)^{-1}\sum_{i=1}^N{\bm
R}_i^{\tau}M_{\hat{\bm F}}({\bm F}\lambda_i+{\bm e}_i).
\end{eqnarray*}
Similar to the proof of Lemma A.9 in \cite{Huang2004}, it is easy to show that
$$\Big\|\Big(\frac{L_{N}}{NT}\sum_{i=1}^N{\bm
R}_i^{\tau}M_{\hat{\bm F}}{\bm R}_i\Big)^{-1}\Big\|_{\infty}\leq C.$$

Next, since $M_{\hat{\bm F}}$ is an idempotent matrix, and invoking Lemma A.6 in \cite{Huang2004} and (\ref{term}),
by a simple calculation, we can obtain that
\begin{eqnarray*}
&&\Big|\frac{L_{N}}{NT}\sum_{i=1}^N{\bm
R}_i^{\tau}M_{\hat{\bm F}}{\bm e}_i\Big|_{\infty}=\Big|\frac{L_{N}}{NT}\sum_{i=1}^N\sum_{t=1}^{T}
R_{it}^{\tau}(M_{\hat{\bm F}}{\bm e}_i)_{t}\Big|_{\infty}\\
&\leq &\max\limits_{k,l}\Big|\frac{L_{N}}{NT}\sum_{i=1}^N\sum_{t=1}^{T}
X_{ik}(u_{it})B_{kl}(u_{it})({\bm e}_{i}^{\tau}M_{\hat{\bm F}}{\bm e}_i)^{1/2}\Big|\\
&\leq &\max\limits_{k,l}\Big|\frac{L_{N}}{NT}\sum_{i=1}^N\sum_{t=1}^{T}
X_{ik}(u_{it})B_{kl}(u_{it})\|{\bm e}_{i}\|\Big|\\
&\leq& L_{N}\max\limits_{k}\sup\limits_{u}|X_{ik}(u)|\max\limits_{k,l}\Big(\frac{1}{NT}\sum_{i=1}^N\sum_{t=1}^{T}
B_{kl}(u_{it})\Big)O_P(\zeta_{Ld}^{1/2})=O_P(\zeta_{Ld}^{1/2}).
\end{eqnarray*}

By $M_{\bm F}{\bm F}=0$ and (\ref{LA6}) in  Appendix B, and Assumptions (A6) and (A7), we have
\begin{eqnarray*}
&&\Big|\frac{L_{N}}{NT}\sum_{i=1}^N{\bm
R}_i^{\tau}M_{\hat{\bm F}}{\bm F}\lambda_i\Big|_{\infty}\\
&=&\Big|\frac{L_{N}}{NT}\sum_{i=1}^N{\bm
R}_i^{\tau}(M_{\hat{\bm F}}-M_{\bm F}){\bm F}\lambda_i\Big|_{\infty}
=\Big|\frac{L_{N}}{NT}\sum_{i=1}^N{\bm
R}_i^{\tau}(P_{\hat{\bm F}}-P_{\bm F}){\bm F}\lambda_i\Big|_{\infty}\\
&=&O_P\Big(L_N^{1/2}\zeta_{Ld}^{1/2}\Big).
\end{eqnarray*}

In addition, by Assumptions (A1) and (A8) (ii), Lemma 1 in  Appendix B, and the properties of B-spline,
similar to the proof of Corollary 1 in \cite{Huang2004},
we can obtain that
\begin{eqnarray*}
&&\varpi_k^{\tau}{\bm B}(u)\left(\sum_{i=1}^N{\bm Z}_i^{\tau}{\bm Z}_i\right)^{-1}\left(\sum_{i=1}^N{\bm Z}_i^{\tau}\Omega_i{\bm Z}_i\right)\left(\sum_{i=1}^N{\bm Z}_i^{\tau}{\bm Z}_i\right)^{-1}{\bm B}(u)^{\tau}\varpi_{k}\\
&\gtrsim &C\frac{L_{N}}{NT}\sum_{l=1}^{L_{k}}B_{kl}^{2}(u)\gtrsim \frac{L_{N}}{NT}.
\end{eqnarray*}
This proves Theorem \ref{theo4}.
\hfill$\Box$

\vskip12pt

\noindent{\bf Proof of Theorem \ref{theo5}.}
Note that
$$({\bf D}^{\tau}{\bf D})^{-1}=\frac{1}{T}\left(I_{N-1}-\frac{1}{N}{\bf 1}_{N-1}{\bf 1}_{N-1}^{\tau}\right),~~
({\bf S}^{\tau}{\bf S})^{-1}=\frac{1}{N}\left(I_{T-1}-\frac{1}{T}{\bf 1}_{T-1}{\bf 1}_{T-1}^{\tau}\right).$$
By a simple calculation, we can get that
$${\bf \Gamma}=I_{NT}-\frac{1}{T}I_{N}\otimes {\bf 1}_{T}{\bf 1}_{T}^{\tau}-\frac{1}{N}{\bf 1}_{N}{\bf 1}_{N}^{\tau}\otimes I_{T}
+\frac{2}{NT}{\bf 1}_{NT}{\bf 1}_{NT}^{\tau}.$$

Hence, similar to the proof of Theorem 2 in \cite{Huang2004} and the proof of Theorem 2.2 in \cite{Ai2014},
applying the standard method, we can show that Theorem \ref{theo5} holds.
\hfill$\Box$

\section*{Appendix B: Some lemmas and their proofs} \label{appB}

\renewcommand{\theequation}{B.\arabic{equation}}
\setcounter{equation}{0}

This appendix  contains  Lemmas \ref{lemB01}--\ref{lemB6} and their proofs.

{\lemma\label{lemB01} Let $\rho_{\min}$ and $\rho_{\max}$ be the
minimum and maximum eigenvalues of $L_ND(\bm F)$ respectively. Then
there exist positive constants $M_3$ and $M_4$ such that
$M_3\le\rho_{\min}\le\rho_{\max}\le M_4$. }

The proof of Lemma \ref{lemB01} follows the same lines as Lemma A.3 in \cite{Huang2004}, Lemma 3.2 in \cite{HeShi1994}, and Lemma 3 in \cite{TangCheng2009}. We hence omit  the proof of Lemma \ref{lemB01}.

{\lemma\label{lemB1} Assume that  assumptions (A1), (A2),
(A4)--(A8) hold. We have
\begin{eqnarray*}
&&\sup_{{\bm F}}\left\|\frac{1}{NT}\sum_{i=1}^N{\bm
R}_i^{\tau}M_{\bm
F}{\bm \varepsilon}_i\right\|=o_P(1),\\
&&\sup_{{\bm F}}\left\|\frac{1}{NT}\sum_{i=1}^N\lambda_i^{\tau}{\bm
F}^{\tau}M_{\hat{\bm F}}{\bm \varepsilon}_i\right\|=o_P(1),\\
&&\sup_{{\bm
F}}\left\|\frac{1}{NT}\sum_{i=1}^N\bm\varepsilon_i^{\tau}P_{{\bm
F}}{\bm \varepsilon}_i\right\|=o_P(1).
\end{eqnarray*}
}

\noindent{\bf Proof.}\ \ Using $P_{\bm F}={\bm F}{\bm F}^{\tau}/T$, we
have
\begin{eqnarray*}
\frac{1}{NT}\sum_{i=1}^N{\bm R}_i^{\tau}M_{\bm F}{\bm
\varepsilon}_i=\frac{1}{NT}\sum_{i=1}^N{\bm R}_i^{\tau}{\bm
\varepsilon}_i-\frac{1}{NT}\sum_{i=1}^N{\bm R}_i^{\tau}P_{\bm F}{\bm
\varepsilon}_i.
\end{eqnarray*}
By Assumptions (A1) and (A8), together with  the properties of B-spline, it is
easy to show that $\frac{1}{NT}\sum_{i=1}^N{\bm R}_i^{\tau}{\bm
\varepsilon}_i=O_P((NT)^{-1/2})=o_P(1)$. Now we  show that
$\sup\limits_{\bm F}\frac{1}{NT}\sum_{i=1}^N{\bm R}_i^{\tau}P_{\bm F}{\bm
\varepsilon}_i=o_P(1)$. Note that
\begin{eqnarray}\nonumber
\frac{1}{NT}\left\|\sum_{i=1}^N{\bm R}_i^{\tau}P_{\bm F}{\bm
\varepsilon}_i\right\|&=&\left\|\frac{1}{N}\sum_{i=1}^N\left(\frac{{\bm
R}_i^{\tau}{\bm
F}}{T}\right)\frac{1}{T}\sum_{t=1}^{T}F_t\varepsilon_{it}\right\|\\\label{ApenB-1}
&\le&\frac{1}{N}\sum_{i=1}^N\left\|\frac{{\bm R}_i^{\tau}{\bm
F}}{T}\right\|\cdot\left\|\frac{1}{T}\sum_{t=1}^{T}F_t\varepsilon_{it}\right\|.
\end{eqnarray}
By $T^{-1/2}\|{\bm F}\|=\sqrt{r}$, we have $T^{-1}\|{\bm
R}_i^{\tau}{\bm F}\|\le T^{-1}\|{\bm R}_i\|\|{\bm
F}\|=\sqrt{r}T^{-1/2}\|{\bm R}_i\|$.  By the Cauchy-Schwarz
inequality, (\ref{ApenB-1}) is bounded above by
\begin{eqnarray*}
\sqrt{r}\left(\frac{1}{N}\sum_{i=1}^N\frac{1}{T}\sum_{t=1}^T\|R_{it}\|^2\right)^{1/2}
\left(\frac{1}{N}\sum_{i=1}^N\left\|\frac{1}{T}\sum_{t=1}^{T}F_t\varepsilon_{it}\right\|^2\right)^{1/2}.
\end{eqnarray*}
By $T^{-1/2}\|{\bm R}_i\|=O_P(1)$, the first term of the above
expression is of order  $O_P(1)$. Similar to the proof of Lemma A.1 in \cite{Bai2009}, it is easy to show that the order of the second term
is $o_P(1)$ uniformly in $\bm F$.
\begin{eqnarray*}
\frac{1}{N}\sum_{i=1}^N\left\|\frac{1}{T}\sum_{t=1}^{T}F_t\varepsilon_{it}\right\|^2
&=&{\rm
tr}\left(\frac{1}{N}\sum_{i=1}^N\frac{1}{T^2}\sum_{t=1}^T\sum_{s=1}^TF_tF_s^{\tau}\varepsilon_{it}\varepsilon_{is}\right)\\
&=&{\rm tr}\left(\frac{1}{N}\sum_{i=1}^N\frac{1}{T^2}
\sum_{t=1}^T\sum_{s=1}^TF_tF_s^{\tau}[\varepsilon_{it}\varepsilon_{is}-E(\varepsilon_{it}\varepsilon_{is})]\right)\\
&&+{\rm tr}\left(\frac{1}{T^2}
\sum_{t=1}^T\sum_{s=1}^TF_tF_s^{\tau}\frac{1}{N}\sum_{i=1}^NE(\varepsilon_{it}\varepsilon_{is})\right).
\end{eqnarray*}
Note that $T^{-1}\sum\limits_{t=1}^T\|F_t\|^2=\|{\bm F}^{\tau}{\bm
F}/T\|=r$. By the Cauchy-Schwarz inequality and Assumption (A8),
we obtain that
\begin{eqnarray*}
&&{\rm tr}\left(\frac{1}{N}\sum_{i=1}^N\frac{1}{T^2}
\sum_{t=1}^T\sum_{s=1}^TF_tF_s^{\tau}[\varepsilon_{it}\varepsilon_{is}-E(\varepsilon_{it}\varepsilon_{is})]\right)\\
&\le& \left(\frac{1}{T^2}
\sum_{t=1}^T\sum_{s=1}^T\|F_t\|^2\|F_s\|^{2}\right)^{1/2}N^{-1/2}
\left(\frac{1}{T^2} \sum_{t=1}^T\sum_{s=1}^T
\left[\frac{1}{\sqrt{N}}\sum_{i=1}^N[\varepsilon_{it}\varepsilon_{is}-E(\varepsilon_{it}\varepsilon_{is})]\right]^2\right)^{1/2}\\
&=&rN^{-1/2}O_P(1).
\end{eqnarray*}
Similarly, we have
\begin{eqnarray*}
&&{\rm tr}\left(\frac{1}{T^2}
\sum_{t=1}^T\sum_{s=1}^TF_tF_s^{\tau}\frac{1}{N}\sum_{i=1}^NE(\varepsilon_{it}\varepsilon_{is})\right)\\
&\le&\left(\frac{1}{T^2}
\sum_{t=1}^T\sum_{s=1}^T\|F_t\|^2\|F_s\|^{2}\right)^{1/2}
 \left(\frac{1}{T^2}
\sum_{t=1}^T\sum_{s=1}^T\left[\frac{1}{N}\sum_{i=1}^NE(\varepsilon_{it}\varepsilon_{is})\right]^2\right)^{1/2}\\
&=&rT^{-1/2}\left(\frac{1}{T}
\sum_{t=1}^T\sum_{s=1}^T\left[\frac{1}{N}\sum_{i=1}^NE(\varepsilon_{it}\varepsilon_{is})\right]^2\right)^{1/2}\\
&=&rO(T^{-1/2}).
\end{eqnarray*}
This shows that
\begin{eqnarray*}
\sup_{{\bm F}}\left\|\frac{1}{NT}\sum_{i=1}^N{\bm R}_i^{\tau}M_{\bm
F}{\bm \varepsilon}_i\right\|=O_P((NT)^{-1/2})=o_P(1).
\end{eqnarray*}
The proofs of the second and third results  are similar to the
proof of the first one, and hence are omitted.  \hfill$\Box$

{\lemma\label{lemB2} Assume that  assumptions  (A1)--(A9) hold. For ease of notation,  let $H=(\Lambda^{\tau}\Lambda/N)({\bm F}^{\tau}\hat{\bm
F}/T)V_{NT}^{-1}$. We have
\begin{eqnarray*}
&({\rm i})& T^{-1/2}\|\hat{\bm F}-{\bm
F}H\|=O_P(\|\hat{\bm\gamma}-\widetilde{\bm\gamma}\|)+O_P(\delta_{NT}^{-1})+O_P\Big(\zeta_{Ld}^{1/2}\Big),\\
&({\rm ii})& T^{-1}{\bm F}^{\tau}(\hat{\bm F}-{\bm
F}H)=O_P(\|\hat{\bm\gamma}-\widetilde{\bm\gamma}\|)+O_P(\delta_{NT}^{-2})+O_P\Big(\zeta_{Ld}^{1/2}\Big),\\
&({\rm iii})& T^{-1}\hat{\bm F}^{\tau}(\hat{\bm F}-{\bm
F}H)=O_P(\|\hat{\bm\gamma}-\widetilde{\bm\gamma}\|)+O_P(\delta_{NT}^{-2})+O_P\Big(\zeta_{Ld}^{1/2}\Big),\\
&({\rm iv})&  T^{-1}{\bm R}_j^{\tau}(\hat{\bm F}-{\bm
F}H)=O_P(\|\hat{\bm\gamma}-\widetilde{\bm\gamma}\|)+O_P(\delta_{NT}^{-2})+O_P\Big(\zeta_{Ld}^{1/2}\Big)~~\hbox{for all $j$},\\
&({\rm v})&  \frac{1}{NT}\sum_{j=1}^N{\bm R}_j^{\tau}M_{\hat{\bm
F}}(\hat{\bm F}-{\bm
F}H)=O_P(\|\hat{\bm\gamma}-\widetilde{\bm\gamma}\|)+O_P(\delta_{NT}^{-2})+O_P\Big(\zeta_{Ld}^{1/2}\Big),\\
&({\rm vi})& HH^{\tau}-(T^{-1}{\bm F}^{\tau}{\bm
F})^{-1}=O_P(\|\hat{\bm\gamma}-\widetilde{\bm\gamma}\|)+O_P(\delta_{NT}^{-2})+O_P\Big(\zeta_{Ld}^{1/2}\Big).
\end{eqnarray*}
}

\noindent{\bf Proof.}\ \ (i)\ \ From (\ref{equation-2}),
(\ref{term}) and ${\bm Y}_i={\bm R}_i\widetilde{\bm\gamma}+{\bm
F}\lambda_i+{\bm \varepsilon}_i+{\bm e}_i$ for $ i=1,\ldots,N$, we have the following expansion:
\begin{eqnarray}\nonumber
\hat{\bm F}V_{NT}&=&\frac{1}{NT}\sum_{i=1}^{N}{\bm
R}_i(\widetilde{\bm\gamma}-\hat{\bm\gamma})(\widetilde{\bm\gamma}-\hat{\bm\gamma})^{\tau}{\bm
R}_i^{\tau}\hat{\bm F}+\frac{1}{NT}\sum_{i=1}^{N}{\bm
R}_i(\widetilde{\bm\gamma}-\hat{\bm\gamma})\lambda_i^{\tau}{\bm
F}^{\tau}\hat{\bm F}\\\nonumber &&+\frac{1}{NT}\sum_{i=1}^{N}{\bm
R}_i(\widetilde{\bm\gamma}-\hat{\bm\gamma}){\bm\varepsilon}_i^{\tau}\hat{\bm
F}+\frac{1}{NT}\sum_{i=1}^{N}{\bm
F}\lambda_i(\widetilde{\bm\gamma}-\hat{\bm\gamma})^{\tau}{\bm
R}_i^{\tau}\hat{\bm F}\\\nonumber &&+\frac{1}{NT}\sum_{i=1}^{N}{\bm
\varepsilon}_i(\widetilde{\bm\gamma}-\hat{\bm\gamma})^{\tau}{\bm
R}_i^{\tau}\hat{\bm F}+\frac{1}{NT}\sum_{i=1}^{N}{\bm
F}\lambda_i{\bm \varepsilon}_i^{\tau}\hat{\bm
F}+\frac{1}{NT}\sum_{i=1}^{N}{\bm \varepsilon}_i\lambda_i^{\tau}{\bm
F}^{\tau}\hat{\bm F}\\\nonumber &&+\frac{1}{NT}\sum_{i=1}^{N}{\bm
\varepsilon}_i{\bm \varepsilon}_i^{\tau}\hat{\bm
F}+\frac{1}{NT}\sum_{i=1}^{N}{\bm
R}_i(\widetilde{\bm\gamma}-\hat{\bm\gamma}){\bm e}_i^{\tau}\hat{\bm
F} +\frac{1}{NT}\sum_{i=1}^{N}{\bm
e}_i(\widetilde{\bm\gamma}-\hat{\bm\gamma})^{\tau}{\bm
R}_i^{\tau}\hat{\bm F}\\\nonumber &&+\frac{1}{NT}\sum_{i=1}^{N}{\bm
F}\lambda_i{\bm e}_i^{\tau}\hat{\bm
F}+\frac{1}{NT}\sum_{i=1}^{N}{\bm e}_i\lambda_i^{\tau}{\bm
F}^{\tau}\hat{\bm F} +\frac{1}{NT}\sum_{i=1}^{N}{\bm
\varepsilon}_i{\bm e}_i^{\tau}\hat{\bm F}\\\nonumber
&&+\frac{1}{NT}\sum_{i=1}^{N}{\bm e}_i{\bm
\varepsilon}_i^{\tau}\hat{\bm F}+\frac{1}{NT}\sum_{i=1}^{N}{\bm
e}_i{\bm e}_i^{\tau}\hat{\bm F}+\frac{1}{NT}\sum_{i=1}^{N}{\bm
F}\lambda_i\lambda_i^{\tau}{\bm F}^{\tau}\hat{\bm F}\\\nonumber
&=:&B_1+B_2+B_3+\cdots+B_{16},
\end{eqnarray}
where $B_{16}=\frac{1}{NT}\sum_{i=1}^{N}{\bm
F}\lambda_i\lambda_i^{\tau}{\bm F}^{\tau}\hat{\bm F}={\bm F}(\Lambda^{\tau}\Lambda/N)({\bm
F}^{\tau}\hat{\bm F}/T)$. This leads to
\begin{eqnarray}\label{A-2-1}
\hat{\bm F}-{\bm F}H=(B_1+B_2+\cdots+B_{15})V_{NT}^{-1}.
\end{eqnarray}
Noting that $T^{-1/2}\|\hat{\bm F}\|=\sqrt{r}$ and
$\|{\bm R}_i\|=O_P(T^{1/2})$, we have
\begin{eqnarray*}
T^{-1/2}\|B_1\|&\le& \frac{1}{N}\sum_{i=1}^N\Big(\frac{\|{\bm
R}_i\|^2}{T}\Big)\|\hat{\bm\gamma}-\widetilde{\bm\gamma}\|^2\sqrt{r}=O_P(\|\hat{\bm\gamma}-\widetilde{\bm\gamma}\|^2)=o_P(\|\hat{\bm\gamma}-\widetilde{\bm\gamma}\|),\\
T^{-1/2}\|B_2\|&\le& \frac{1}{N}\sum_{i=1}^N\Big(\frac{\|{\bm
R}_i\|}{\sqrt{T}}\Big)\|\hat{\bm\gamma}-\widetilde{\bm\gamma}\|\|\lambda_i\|\|{\bm
F}^{\tau}\hat{{\bm
F}}/T\|=O_P(\|\hat{\bm\gamma}-\widetilde{\bm\gamma}\|).
\end{eqnarray*}
Using the same argument, it is easy to show that
$T^{-1/2}\|B_l\|=O_P(\|\hat{\bm\gamma}-\widetilde{\bm\gamma}\|)$ for
$l=3,4$ and 5, and $T^{-1/2}\|B_l\|=O_P(\delta_{NT}^{-1})$ for $l=6,7$
and 8. For $B_9$, using the same argument, and by (\ref{term}) and
Assumption (A1), we have
\begin{eqnarray*}
T^{-1/2}\|B_9\|&\le&
T^{-1/2}\frac{1}{N}\sum_{i=1}^N\Big(\frac{\|{\bm
R}_i\|}{\sqrt{T}}\Big)\|\hat{\bm\gamma}-\widetilde{\bm\gamma}\|\Big(\frac{\|\hat{\bm F}\|}{\sqrt{T}}\Big)\sqrt{\sum_{t=1}^Te_{it}^2}\\
&\le&O_P(\|\hat{\bm\gamma}-\widetilde{\bm\gamma}\|)\cdot
M\zeta_{Ld}^{1/2}.
\end{eqnarray*}
Similarly, we can prove that
$T^{-1/2}\|B_{10}\|=O_P(\|\hat{\bm\gamma}-\widetilde{\bm\gamma}\|)\cdot
M\zeta_{Ld}^{1/2}.$ For $B_{11}$, we have
\begin{eqnarray*}
T^{-1/2}\|B_{11}\|\le
T^{-1/2}\frac{1}{N}\sum_{i=1}^N\Big(\frac{\|{\bm
F}\|}{\sqrt{T}}\Big)\|\lambda_i\|\sqrt{r\sum_{t=1}^Te_{it}^2}
=O_P\Big(\zeta_{Ld}^{1/2}\Big).
\end{eqnarray*}
Similarly, it yields that
$T^{-1/2}\|B_{12}\|=O_P(\zeta_{Ld}^{1/2}).$
For $B_{13}$, we have
\begin{eqnarray*}
T^{-1/2}\|B_{13}\|\le \frac{1}{NT}\sum_{i=1}^N\|{\bm
\varepsilon}_i\|\sqrt{r\sum_{t=1}^Te_{it}^2}
=O_P\Big(\zeta_{Ld}^{1/2}\delta_{NT}^{-1}\Big).
\end{eqnarray*}
Similarly, it yields that
$T^{-1/2}\|B_{14}\|=O_P(\zeta_{Ld}^{1/2}\delta_{NT}^{-1}).$
For $B_{15}$, we have
\begin{eqnarray*}
T^{-1/2}\|B_{15}\|\le
\frac{1}{NT}\sum_{i=1}^N\Big(\sum_{t=1}^Te_{it}^2\Big)\sqrt{r}
=O_P(\zeta_{Ld}).
\end{eqnarray*}
Following the same arguments as in the proof of Proposition A.1
in \cite{Bai2009}, together with the above results,  we have
\begin{eqnarray*}
T^{-1/2}\|\hat{\bm F}-{\bm
F}H\|=O_P(\|\hat{\bm\gamma}-\widetilde{\bm\gamma}\|)+O_P(\delta_{NT}^{-1})+O_P\Big(\zeta_{Ld}^{1/2}\Big).
\end{eqnarray*}

(ii)\ \ By (\ref{A-2-1}), we have the following decomposition:
\begin{eqnarray*}\label{ApenB-3}
T^{-1}{\bm F}^{\tau}(\hat{\bm F}-{\bm F}H)=T^{-1}{\bm
F}^{\tau}(B_1+B_2+\cdots+B_{15})V_{NT}^{-1}.
\end{eqnarray*}
Invoking the similar arguments as in the proof of Lemma A.3 (i) in \cite{Bai2009s} to the first eight terms, we can obtain that
\begin{eqnarray*}
T^{-1}{\bm
F}^{\tau}(B_1+B_2+\cdots+B_{8})V_{NT}^{-1}=O_P(\|\hat{\bm\gamma}-\widetilde{\bm\gamma}\|)+O_P(\delta_{NT}^{-2}).
\end{eqnarray*}
For the other terms, we can   show that $T^{-1}{\bm F}^{\tau}B_9V_{NT}^{-1}$ and
$T^{-1}{\bm F}^{\tau}B_{10}V_{NT}^{-1}$ are of order
$O_P(\|\hat{\bm\gamma}-\widetilde{\bm\gamma}\|\zeta_{Ld}^{1/2}),$
$T^{-1}{\bm F}^{\tau}B_{11}V_{NT}^{-1}$ and $T^{-1}{\bm
F}^{\tau}B_{12}V_{NT}^{-1}$ are of order
$O_P(\zeta_{Ld}^{1/2})$, $T^{-1}{\bm
F}^{\tau}B_{13}V_{NT}^{-1}$ and $T^{-1}{\bm
F}^{\tau}B_{14}V_{NT}^{-1}$ are of order
$O_P(\zeta_{Ld}^{1/2}\delta_{NT}^{-1})$,
and $T^{-1}{\bm
F}^{\tau}B_{15}V_{NT}^{-1}=O_P(\zeta_{Ld}).$
This finishes the proof of (ii).

(iii)\ \ By (i) and (ii) and some elementary calculations, we have
\begin{eqnarray*}
\|T^{-1}\hat{\bm F}^{\tau}(\hat{\bm F}-{\bm F}H)\|&\le&
T^{-1}\|\hat{\bm F}-{\bm F}H\|^2+\|H\|T^{-1}\|{\bm
F}^{\tau}(\hat{\bm F}-{\bm F}H)\|\\
&=&O_P(\|\hat{\bm\gamma}-\widetilde{\bm\gamma}\|)+O_P(\delta_{NT}^{-2})+O_P\Big(\zeta_{Ld}^{1/2}\Big).
\end{eqnarray*}

(iv)\ \  The proof of (iv) is similar to that for  (ii), and hence is omitted.

(v)\ \ Noting that $M_{\hat{\bm F}}=I_T-\hat{\bm F}\hat{\bm
F}^{\tau}/T$, we have
\begin{eqnarray*}
 &&\frac{1}{NT}\sum_{j=1}^N{\bm R}_j^{\tau}M_{\hat{\bm
F}}(\hat{\bm F}-{\bm F}H)\\
&=&\frac{1}{N}\sum_{j=1}^N\frac{1}{T}{\bm
R}_j^{\tau}(\hat{\bm F}-{\bm F}H)-\frac{1}{N}\sum_{j=1}^N\frac{{\bm
R}_j^{\tau}\hat{\bm F}}{T}T^{-1}\hat{\bm F}^{\tau}(\hat{\bm F}-{\bm F}H)\\
&=:&I_1+I_2.
\end{eqnarray*}
Since $I_1$ is an average of $\frac{1}{T}{\bm R}_j^{\tau}(\hat{\bm
F}-{\bm F}H)$ over $j$,  it is easy to verify  that
$I_1=O_P(\|\hat{\bm\gamma}-\widetilde{\bm\gamma}\|)+O_P(\delta_{NT}^{-2})+O_P(\zeta_{Ld}^{1/2})$.
For $I_2$, by (iii) we have
\begin{eqnarray*}
\|I_2\|&\le& \frac{1}{N}\sum_{j=1}^N\frac{\|{\bm
R}_j\|}{\sqrt{T}}\sqrt{r}\|T^{-1}\hat{\bm F}^{\tau}(\hat{\bm F}-{\bm
F}H)\|\\
&=&O_P(\|\hat{\bm\gamma}-\widetilde{\bm\gamma}\|)+O_P(\delta_{NT}^{-2})+O_P\Big(\zeta_{Ld}^{1/2}\Big).
\end{eqnarray*}
This completes  the proof of (v).

(vi)\ \ By (ii), we have
\begin{eqnarray}\label{ApendB-4}
{\bm F}^{\tau}\hat{\bm F}/T-({\bm F}^{\tau}{\bm
F}/T)H=O_P(\|\hat{\bm\gamma}-\widetilde{\bm\gamma}\|)+O_P(\delta_{NT}^{-2})+O_P\Big(\zeta_{Ld}^{1/2}\Big).
\end{eqnarray}
By (iii) and the fact that $\hat{\bm F}^{\tau}\hat{\bm F}/T=I_r$, we have
\begin{eqnarray}\label{ApendB-5}
I_r-(\hat{\bm F}^{\tau}{\bm
F}/T)H=O_P(\|\hat{\bm\gamma}-\widetilde{\bm\gamma}\|)+O_P(\delta_{NT}^{-2})+O_P\Big(\zeta_{Ld}^{1/2}\Big).
\end{eqnarray}
Left-multiplying by $H^{\tau}$ in (\ref{ApendB-4}), and using the
transpose for (\ref{ApendB-5}), we have
\begin{eqnarray*}
I_r-H^{\tau}({\bm F}^{\tau}{\bm
F}/T)H=O_P(\|\hat{\bm\gamma}-\widetilde{\bm\gamma}\|)+O_P(\delta_{NT}^{-2})+O_P\Big(\zeta_{Ld}^{1/2}\Big),
\end{eqnarray*}
which shows that (vi) holds.   \hfill$\Box$

{\lemma\label{lemB3} Assume that  assumptions (A1)-- (A9) hold.
We have
\begin{eqnarray*}
&({\rm i})& T^{-1}{\bm \varepsilon}_j^{\tau}(\hat{\bm F}-{\bm
F}H)=T^{-1/2}O_P(\|\hat{\bm\gamma}-\widetilde{\bm\gamma}\|)+O_P(\delta_{NT}^{-2})\\
&&~~~~~~~~~~~~~+O_P\Big(\zeta_{Ld}^{1/2}T^{-1/2}\Big)
~~ \hbox{for all $j=1,\ldots,N$},\\
&({\rm ii})& \frac{1}{T\sqrt{N}}\sum_{j=1}^N {\bm
\varepsilon}_j^{\tau}(\hat{\bm F}-{\bm
F}H)=T^{-1/2}O_P(\|\hat{\bm\gamma}-\widetilde{\bm\gamma}\|)
+N^{-1/2}O_P(\|\hat{\bm\gamma}-\widetilde{\bm\gamma}\|)\\
&&~~~~~~~~~~~~~+O_P(N^{-1/2})+O_P(\delta_{NT}^{-2})+O_P\Big(\zeta_{Ld}^{1/2}\Big),\\
&({\rm iii})& \frac{1}{NT}\sum_{j=1}^N\lambda_j {\bm
\varepsilon}_j^{\tau}(\hat{\bm F}-{\bm
F}H)=(TN)^{-1/2}O_P(\|\hat{\bm\gamma}-\widetilde{\bm\gamma}\|) +
O_P(N^{-1})\\
&&~~~~~~~~~~~~~+N^{-1/2}O_P(\delta_{NT}^{-2})+N^{-1/2}O_P\Big(\zeta_{Ld}^{1/2}\Big).
\end{eqnarray*}
}

\noindent{\bf Proof.}\ \ (i)\ \ By (\ref{A-2-1}), we have
\begin{eqnarray}\label{ApenB-2}
T^{-1}{\bm \varepsilon}_j^{\tau}(\hat{\bm F}-{\bm F}H)=T^{-1}{\bm
\varepsilon}_j^{\tau}(B_1+B_2+\cdots+B_{15})V_{NT}^{-1}.
\end{eqnarray}
Invoking the similar arguments as in the proof of Lemma A.4 (i) in \cite{Bai2009s} to the first eight terms, we can obtain that
\begin{eqnarray*}
T^{-1}{\bm
\varepsilon}_j^{\tau}(B_1+B_2+\cdots+B_{8})V_{NT}^{-1}=T^{-1/2}O_P(\|\hat{\bm\gamma}-\widetilde{\bm\gamma}\|)+O_P(\delta_{NT}^{-2}).
\end{eqnarray*}
For the other terms in (\ref{ApenB-2}), similar to the proof of (i)
in Lemma \ref{lemB2}, we only need to show that the dominant terms
$T^{-1}{\bm \varepsilon}_j^{\tau}B_{11}V_{NT}^{-1}$ and $T^{-1}{\bm
\varepsilon}_j^{\tau}B_{12}V_{NT}^{-1}$ are the same order as
$O_P(\zeta_{Ld}^{1/2}T^{-1/2})$. For
$T^{-1}{\bm \varepsilon}_j^{\tau}B_{11}V_{NT}^{-1}$, we have
\begin{small}
\begin{eqnarray*}
\Big\|T^{-1}{\bm \varepsilon}_j^{\tau}B_{11}V_{NT}^{-1}\Big\|\le
\frac{1}{\sqrt{T}}\frac{\|{\bm\varepsilon}_j^{\tau}{\bm
F}\|}{\sqrt{T}}\frac{1}{N\sqrt{T}}\sum_{i=1}^N\|\lambda_i\|\|V_{NT}^{-1}\|\sqrt{r\sum_{t=1}^Te_{it}^2}
=O_P\Big(\zeta_{Ld}^{1/2}T^{-1/2}\Big).
\end{eqnarray*}
\end{small}
This leads to $T^{-1/2}\|{\bm\varepsilon}_j^{\tau}{\bm F}\|=O_P(1)$. Similarly,
$\|T^{-1}{\bm
\varepsilon}_j^{\tau}B_{12}V_{NT}^{-1}\Big\|=O_P(\zeta_{Ld}^{1/2}T^{-1/2})$.
Thus, we finish the proof of (i).

(ii)\ \ By (\ref{A-2}), we have
\begin{eqnarray*}
\frac{1}{T\sqrt{N}}\sum_{j=1}^N {\bm \varepsilon}_j^{\tau}(\hat{\bm
F}H^{-1}-{\bm F})&=&\frac{1}{T\sqrt{N}}\sum_{j=1}^N {\bm
\varepsilon}_j^{\tau}(B_1+B_2+\cdots+B_{15})G\\
&=:&a_1+\cdots+a_{15}.
\end{eqnarray*}
Next we derive the orders of the fifteen terms, respectively.
For the first four terms, we have
\begin{eqnarray*}
\|a_1\|&\le& T^{-1/2}\|G\|\left(\frac{1}{N}\sum_{i=1}^N
\Big\|\frac{1}{\sqrt{NT}}\sum_{j=1}^N
\sum_{t=1}^T\varepsilon_{jt}R_{it}\Big\|\Big(\frac{\|{\bm
R}_i\|^2}{T}\Big)\right)\|\hat{\bm\gamma}-\widetilde{\bm\gamma}\|^2\\
&=&T^{-1/2}O_P(\|\hat{\bm\gamma}-\widetilde{\bm\gamma}\|^2),\\
a_2&=&\frac{1}{NT}\frac{1}{\sqrt{N}}\sum_{j=1}^N
\sum_{i=1}^N{\bm\varepsilon}_j^{\tau}{\bm
R}_i(\widetilde{\bm\gamma}-\hat{\bm\gamma})\lambda_i^{\tau}\Big(\frac{\Lambda^{\tau}\Lambda}{N}\Big)^{-1}\\
&=&\frac{1}{\sqrt{T}}\frac{1}{N}\sum_{i=1}^N\frac{1}{\sqrt{NT}}\sum_{j=1}^N
\sum_{t=1}^TR_{it}\varepsilon_{jt}(\widetilde{\bm\gamma}-\hat{\bm\gamma})\lambda_i^{\tau}\Big(\frac{\Lambda^{\tau}\Lambda}{N}\Big)^{-1}\\
&=&T^{-1/2}O_P(\|\hat{\bm\gamma}-\widetilde{\bm\gamma}\|),\\
\|a_3\|&\le&T^{-1/2}\|G\|\left(\frac{1}{N}\sum_{i=1}^N
\Big\|\frac{1}{\sqrt{NT}}\sum_{j=1}^N
\sum_{t=1}^T\varepsilon_{jt}R_{it}\Big\|\Big(\frac{\|{\bm
\varepsilon}_i\|^2}{T}\Big)\right)\|\hat{\bm\gamma}-\widetilde{\bm\gamma}\|\\
&=&T^{-1/2}O_P(\|\hat{\bm\gamma}-\widetilde{\bm\gamma}\|),\\
\|a_4\|&\le&T^{-1/2}\|G\|\left\|\frac{1}{\sqrt{NT}}\sum_{j=1}^N
\sum_{t=1}^T\varepsilon_{jt}F_t^{\tau}\right\|
\left\|\frac{1}{N}\sum_{i=1}^N\Big(\frac{{\bm R}_i^{\tau}\hat{\bm F}}{T}\Big)\right\|\|\lambda_i\|\|\hat{\bm\gamma}-\widetilde{\bm\gamma}\|\\
&=&T^{-1/2}O_P(\|\hat{\bm\gamma}-\widetilde{\bm\gamma}\|).
\end{eqnarray*}
For $a_5$, let ${\bm W}_i={\bm R}_i^{\tau}\hat{\bm F}/T$. It is
easy to verify  that $\|{\bm W}_i\|^2\le \|{\bm R}_i\|^2/T=O_P(1)$. Further,
\begin{eqnarray*}
a_5&=&\frac{1}{NT}\frac{1}{\sqrt{N}}\sum_{j=1}^N \sum_{t=1}^T{\bm
\varepsilon}_j^{\tau}{\bm
\varepsilon}_i(\widetilde{\bm\gamma}-\hat{\bm\gamma})^{\tau}{\bm
W}_iG\\
&=&\frac{1}{\sqrt{N}}\frac{1}{T}\sum_{t=1}^T\Big(\frac{1}{\sqrt{N}}\sum_{j=1}^N\varepsilon_{jt}\Big)
\Big(\frac{1}{\sqrt{N}}\sum_{i=1}^N\varepsilon_{it}(\widetilde{\bm\gamma}-\hat{\bm\gamma})^{\tau}{\bm
W}_i\Big)G\\
&=&N^{-1/2}O_P(\|\hat{\bm\gamma}-\widetilde{\bm\gamma}\|).
\end{eqnarray*}
For $a_6$, we have
\begin{eqnarray*}
a_6&=&\frac{1}{NT^2}\frac{1}{\sqrt{N}}\sum_{j=1}^N{\bm
\varepsilon}_j^{\tau}{\bm
F}\sum_{i=1}^N\lambda_i{\bm\varepsilon}_i^{\tau}\hat{\bm F}G\\
&=&\frac{1}{NT^2}\frac{1}{\sqrt{N}}\sum_{j=1}^N{\bm
\varepsilon}_j^{\tau}{\bm
F}\sum_{i=1}^N\lambda_i{\bm\varepsilon}_i^{\tau}{\bm
F}HG+\frac{1}{NT^2}\frac{1}{\sqrt{N}}\sum_{j=1}^N{\bm
\varepsilon}_j^{\tau}{\bm
F}\sum_{i=1}^N\lambda_i{\bm\varepsilon}_i^{\tau}(\hat{\bm F}-{\bm
F}H)G\\
&=:&a_{6.1}+a_{6.2}.
\end{eqnarray*}
By the proof of Lemma A.4 in \cite{Bai2009s},
$a_{6.1}=O_P(T^{-1}N^{-1/2})$. Also,
\begin{eqnarray*}
a_{6.2}=T^{-1/2}\left(\frac{1}{\sqrt{NT}}\sum_{j=1}^N\sum_{t=1}^T\varepsilon_{jt}F_t^{\tau}\right)
\frac{1}{NT}\sum_{i=1}^N\lambda_i{\bm\varepsilon}_i^{\tau}(\hat{\bm
F}-{\bm F}H)G.
\end{eqnarray*}
By (i) of Lemma \ref{lemB2} and some elementary calculations, we
have
\begin{eqnarray*}
\|a_{6.2}\|&\le&
T^{-1/2}O_P(1)\frac{1}{N}\sum_{i=1}^N\|\lambda_i\|\|T^{-1/2}{\bm\varepsilon}_i\|
\frac{\|\hat{\bm F}-{\bm F}H\|}{\sqrt{T}}\|G\|\\
&=&T^{-1/2}\Big[O_P(\|\hat{\bm\gamma}-\widetilde{\bm\gamma}\|)+O_P(\delta_{NT}^{-1})+O_P\Big(\zeta_{Ld}^{1/2}\Big)\Big].
\end{eqnarray*}
Since $a_7$ and $a_8$ have the same structures as $a_7$ and $a_8$ in
\cite{Bai2009s}, we can prove that $a_7=O_P(N^{-1/2})$ and
$a_8=O_P(T^{-1})+O_P((NT)^{-1/2})+N^{-1/2}[O_P(\|\hat{\bm\gamma}-\widetilde{\bm\gamma}\|)
+O_P(\delta_{NT}^{-1})+O_P(\zeta_{Ld}^{1/2})].$
For $a_9$, by (\ref{term}) we have
\begin{eqnarray*}
\|a_9\|&\le&\frac{1}{\sqrt{T}}\frac{1}{N}\sum_{i=1}^N\left\|\frac{1}{\sqrt{NT}}\sum_{j=1}^N
\sum_{t=1}^T\varepsilon_{jt}R_{it}\right\|T^{-1/2}\sqrt{r\sum_{t=1}^Te_{it}^2}\|\hat{\bm\gamma}-\widetilde{\bm\gamma}\|\|G\|\\
&=&T^{-1/2}O_P\Big(\|\hat{\bm\gamma}-\widetilde{\bm\gamma}\|\zeta_{Ld}^{1/2}\Big).
\end{eqnarray*}
Similarly,
$a_{10}=T^{-1/2}O_P(\|\hat{\bm\gamma}-\widetilde{\bm\gamma}\|\zeta_{Ld}^{1/2}).$
For $a_{11}$, we have
\begin{eqnarray*}
\|a_{11}\|&\le&T^{-1/2}\left\|\frac{1}{\sqrt{NT}}\sum_{j=1}^N\sum_{t=1}^T\varepsilon_{jt}F_t^{\tau}\right\|
\frac{1}{N}\sum_{i=1}^N\|\lambda_i\|T^{-1/2}\sqrt{r\sum_{t=1}^Te_{it}^2}\|G\|\\
&=&T^{-1/2}O_P\Big(\zeta_{Ld}^{1/2}\Big).
\end{eqnarray*}
For $a_{12}$, we have
\begin{eqnarray*}
a_{12}&=&\frac{1}{\sqrt{N}}\frac{1}{NT}\sum_{j=1}^N\sum_{i=1}^N{\bm\varepsilon}_j^{\tau}{\bm
e}_i\lambda_i^{\tau}\Big(\frac{\Lambda^{\tau}\Lambda}{N}\Big)^{-1}\\
&=&\frac{1}{T}\sum_{t=1}^T
\left[\left(\frac{1}{\sqrt{N}}\sum_{j=1}^N\varepsilon_{jt}\right)
\left(\frac{1}{N}\sum_{i=1}^Ne_{it}\lambda_i^{\tau}\right)\right]\Big(\frac{\Lambda^{\tau}\Lambda}{N}\Big)^{-1}\\
&=&O_P\Big(\zeta_{Ld}^{1/2}\Big).
\end{eqnarray*}
For $a_{13}$, let $\widetilde{\bm W}_i={\bm e}_i^{\tau}\hat{\bm
F}/T$. Then we have $\|\widetilde{\bm W}_i\|=\|{\bm
e}_i\|\sqrt{r}/\sqrt{T}=O_P(\zeta_{Ld}^{1/2})$ and
\begin{eqnarray*}
a_{13}&=&\frac{1}{\sqrt{N}}\frac{1}{T}\sum_{t=1}^T
\left[\left(\frac{1}{\sqrt{N}}\sum_{j=1}^N\varepsilon_{jt}\right)
\left(\frac{1}{\sqrt{N}}\sum_{i=1}^N\varepsilon_{it}\widetilde{{\bm
W}}_i\right)\right]G\\
&=&N^{-1/2}O_P\Big(\zeta_{Ld}^{1/2}\Big).
\end{eqnarray*}
Finally,  we can obtain that
\begin{eqnarray*}
a_{14}=N^{-1/2}O_P\Big(\zeta_{Ld}^{1/2}\Big)~~
\hbox{and}~~a_{15}=O_P(\zeta_{Ld}).
\end{eqnarray*}
Summarizing the above results, we finish the proof of (ii).

(iii)\ \ Part (iii) follows immediately  from (ii) by noting that
$N^{1/2}$  is a constant and that the  presence of $\lambda_k$ does not alter the results.
  \hfill$\Box$

{\lemma\label{lemB4} Assume that  assumptions (A1)-- (A9) hold. We have
\begin{eqnarray*}
&&\frac{1}{N^2T^2}\sum_{i=1}^N\sum_{j=1}^N{\bm
R}_i^{\tau}M_{\hat{\bm F}}({\bm \varepsilon}_j {\bm
\varepsilon}_j^{\tau}-\Omega_j)\hat{\bm F}G\lambda_i\\
&=&O_P(1/(T\sqrt{N}))+(NT)^{-1/2}\left[O_P(\|\hat{\bm\gamma}-\widetilde{\bm\gamma}\|)+O_P(\delta_{NT}^{-1})+O_P\Big(\zeta_{Ld}^{1/2}\Big)\right]\\
&&+\frac{1}{\sqrt{N}}\left[O_P(\|\hat{\bm\gamma}-\widetilde{\bm\gamma}\|)+O_P(\delta_{NT}^{-1})+O_P\Big(\zeta_{Ld}^{1/2}\Big)\right]^2.
\end{eqnarray*}
}

\noindent{\bf Proof.}\ \ Some elementary calculations yield that
\begin{eqnarray*}
&&\frac{1}{N^2T^2}\sum_{i=1}^N\sum_{j=1}^N{\bm
R}_i^{\tau}M_{\hat{\bm F}}({\bm \varepsilon}_j {\bm
\varepsilon}_j^{\tau}-\Omega_j)\hat{\bm F}G\lambda_i\\
&=&\frac{1}{N^2T^2}\sum_{i=1}^N\sum_{j=1}^N{\bm R}_i^{\tau}({\bm
\varepsilon}_j {\bm \varepsilon}_j^{\tau}-\Omega_j)\hat{\bm
F}G\lambda_i\\
&&-\frac{1}{N^2T^2}\sum_{i=1}^N\sum_{j=1}^N{\bm
R}_i^{\tau}\Big(\frac{\hat{\bm F}\hat{\bm F}^{\tau}}{T}\Big)({\bm
\varepsilon}_j {\bm \varepsilon}_j^{\tau}-\Omega_j)\hat{\bm
F}G\lambda_i\\
&=:&I+II.
\end{eqnarray*}
For the first term, by some basic calculations we have
\begin{eqnarray*}
I&=&\frac{1}{N^2T^2}\sum_{i=1}^N\sum_{j=1}^N{\bm R}_i^{\tau}({\bm
\varepsilon}_j {\bm \varepsilon}_j^{\tau}-\Omega_j){\bm
F}HG\lambda_i\\
&&+\frac{1}{N^2T^2}\sum_{i=1}^N\sum_{j=1}^N{\bm R}_i^{\tau}({\bm
\varepsilon}_j {\bm \varepsilon}_j^{\tau}-\Omega_j)(\hat{\bm
F}-{\bm F}H)G\lambda_i\\
&=:&I_1+I_2.
\end{eqnarray*}
Invoking Lemma A.2 (i) in \cite{Bai2009}, we have
$I_1=O_P(1/(T\sqrt{N}))$. Let
\begin{eqnarray*}
a_s=\frac{1}{\sqrt{NT}}\sum_{j=1}^N\sum_{t=1}^T
R_{it}[\varepsilon_{jt}\varepsilon_{js}-E(\varepsilon_{jt}\varepsilon_{js})]=O_P(1).
\end{eqnarray*}
Then we have
\begin{eqnarray*}
I_2=\frac{1}{\sqrt{NT}}\frac{1}{N}\sum_{i=1}^N\frac{1}{T}\sum_{s=1}^Ta_s(\hat{
F}_s-{ F}_sH)^{\tau}G\lambda_i.
\end{eqnarray*}
By the Cauchy-Schwarz inequality and Lemma \ref{lemB2} (i), we have
\begin{eqnarray*}
\Big\|\frac{1}{T}\sum_{s=1}^Ta_s(\hat{ F}_s-{ F}_sH)\Big\|&\le&
\Big(\frac{1}{T}\sum_{s=1}^T\|a_s\|^2\Big)^{1/2}\Big(\frac{1}{T}\sum_{s=1}^T\|\hat{
F}_s-{ F}_sH\|^2\Big)^{1/2}\\
&=&O_P(\|\hat{\bm\gamma}-\widetilde{\bm\gamma}\|)+O_P(\delta_{NT}^{-1})+O_P\Big(\zeta_{Ld}^{1/2}\Big).
\end{eqnarray*}
This leads to
$$I_2=(NT)^{-1/2}\left[O_P(\|\hat{\bm\gamma}-\widetilde{\bm\gamma}\|)+O_P(\delta_{NT}^{-1})+O_P\Big(\zeta_{Ld}^{1/2}\Big)\right].$$

For the second term, by the similar proof of Lemma A.2
(ii) in Bai (2009), we have
\begin{eqnarray*}
\|II\|&\le&\frac{1}{N}\sum_{i=1}^N\Big\|\frac{{\bm
R}_i^{\tau}\hat{\bm
F}}{T}\Big\|\|G\lambda_i\|\Big\|\frac{1}{NT^2}\sum_{j=1}^N\hat{\bm
F}^{\tau}({\bm \varepsilon}_j {\bm
\varepsilon}_j^{\tau}-\Omega_j)\hat{\bm F}\Big\|\\
&=&O_P(1)\Big\|\frac{1}{NT^2}\sum_{j=1}^N\hat{\bm F}^{\tau}({\bm
\varepsilon}_j {\bm \varepsilon}_j^{\tau}-\Omega_j)\hat{\bm
F}\Big\|\\
&=&O_P(1/(T\sqrt{N}))+(NT)^{-1/2}\left[O_P(\|\hat{\bm\gamma}-\widetilde{\bm\gamma}\|)+O_P(\delta_{NT}^{-1})+O_P\Big(\zeta_{Ld}^{1/2}\Big)\right]\\
&&+\frac{1}{\sqrt{N}}\left[O_P(\|\hat{\bm\gamma}-\widetilde{\bm\gamma}\|)+O_P(\delta_{NT}^{-1})+O_P\Big(\zeta_{Ld}^{1/2}\Big)\right]^2.
\end{eqnarray*}
Summarizing the above results, we finish the proof of Lemma
\ref{lemB4}. \hfill$\Box$

{\lemma\label{lemB5} Assume that  assumptions (A1)-- (A9) hold. We have
\begin{eqnarray*}
&&\frac{1}{NT}\sum_{i=1}^N\left[{\bm R}_i^{\tau}M_{\hat{\bm
F}}-\frac{1}{N}\sum_{j=1}^Na_{ij}{\bm R}_j^{\tau}M_{\hat{\bm
F}}\right]{\bm\varepsilon}_i\\
&=&\frac{1}{NT}\sum_{i=1}^N\left[{\bm R}_i^{\tau}M_{{\bm
F}}-\frac{1}{N}\sum_{j=1}^Na_{ij}{\bm R}_j^{\tau}M_{{\bm
F}}\right]{\bm\varepsilon}_i
+N^{-1}\xi_{NT}^*+N^{-1/2}O_P(\|\hat{\bm\gamma}-\widetilde{\bm\gamma}\|^2)\\
&&+(NT)^{-1/2}O_P(\|\hat{\bm\gamma}-\widetilde{\bm\gamma}\|)+N^{-1/2}O_P(\delta_{NT}^{-2})+N^{-1/2}O_P\Big(\zeta_{Ld}^{1/2}\Big),
\end{eqnarray*}
where
\begin{eqnarray*}
\xi_{NT}^*=-\frac{1}{N}\sum_{i=1}^N\sum_{j=1}^N\frac{({\bm R}_i-{\bm
V}_i)^{\tau}{\bm F}}{T}\left(\frac{{\bm F}^{\tau}{\bm
F}}{T}\right)^{-1}\left(\frac{\Lambda^{\tau}\Lambda}{N}\right)^{-1}\lambda_j
\left(\frac{1}{T}\sum_{t=1}^T\varepsilon_{it}\varepsilon_{jt}\right)=O_P(1)
\end{eqnarray*}
with ${\bm V}_i=N^{-1}\sum\limits_{j=1}^Na_{ij}{\bm R}_j$. }

\noindent{\bf Proof.}\ \ For the term
$\dfrac{1}{NT}\sum\limits_{i=1}^N{\bm R}_i^{\tau}(M_{{\bm F}}-M_{\hat{\bm
F}}){\bm\varepsilon}_i$, we consider the following decomposition:
\begin{eqnarray*}\nonumber
\hskip-0.5cm M_{\bm F}-M_{\hat{\bm F}}&=&P_{\hat{\bm F}}-P_{\bm F}\\\nonumber
&=&T^{-1}(\hat{\bm F}-{\bm F}H)H^{\tau}{\bm
F}^{\tau}+T^{-1}(\hat{\bm F}-{\bm F}H)(\hat{\bm F}-{\bm
F}H)^{\tau}\\\nonumber &&+T^{-1}{\bm F}H(\hat{\bm F}-{\bm
F}H)^{\tau}\\
&&+T^{-1}{\bm
F}[HH^{\tau}-(T^{-1}{\bm F}^{\tau}{\bm F})^{-1}]{\bm F}^{\tau}
\end{eqnarray*}
for any invertible matrix $H$. Therefore, we have
\begin{eqnarray*}
&&\frac{1}{NT}\sum_{i=1}^N{\bm R}_i^{\tau}(M_{{\bm F}}-M_{\hat{\bm
F}}){\bm\varepsilon}_i\\
&=&\frac{1}{NT}\sum_{i=1}^N\frac{{\bm R}_i^{\tau}(\hat{\bm F}-{\bm
F}H)}{T}H^{\tau}{\bm
F}^{\tau}{\bm\varepsilon}_i+\frac{1}{NT}\sum_{i=1}^N\frac{{\bm
R}_i^{\tau}(\hat{\bm F}-{\bm F}H)}{T}(\hat{\bm F}-{\bm
F}H)^{\tau}{\bm\varepsilon}_i\\
&&+\frac{1}{NT}\sum_{i=1}^N\frac{{\bm R}_i^{\tau}{\bm
F}H}{T}(\hat{\bm F}-{\bm
F}H)^{\tau}{\bm\varepsilon}_i+\frac{1}{NT}\sum_{i=1}^N\frac{{\bm
R}_i^{\tau}{\bm F}}{T}[HH^{\tau}-(T^{-1}{\bm F}^{\tau}{\bm
F})^{-1}]{\bm F}^{\tau}{\bm\varepsilon}_i\\
&=:&s_1+s_2+s_3+s_4.
\end{eqnarray*}
For $s_1$, noting that
$(\hat{F}_s-H^{\tau}F_s)^{\tau}H^{\tau}F_t$ is scalar,  we have
\begin{eqnarray*}
s_1=\frac{1}{\sqrt{NT}}\frac{1}{T}\sum_{s=1}^T(\hat{F}_s-H^{\tau}F_s)^{\tau}H^{\tau}\left(\frac{1}{\sqrt{NT}}
\sum_{i=1}^N\sum_{t=1}^TF_tR_{is}\varepsilon_{it}\right).
\end{eqnarray*}
Further, we can derive that
\begin{eqnarray*}
\|s_1\|&\le&
\frac{1}{\sqrt{NT}}\left[\frac{1}{T}\sum_{s=1}^T\|\hat{F}_s-H^{\tau}F_s\|^{2}\right]^{1/2}\|H\|
\left[\frac{1}{T}\sum_{s=1}^T\left\|\frac{1}{\sqrt{NT}}
\sum_{i=1}^N\sum_{t=1}^TF_tR_{is}\varepsilon_{it}\right\|^2\right]^{1/2}\\
&=&\frac{1}{\sqrt{NT}}\left[O_P(\|\hat{\bm\gamma}-\widetilde{\bm\gamma}\|)
+O_P(\delta_{NT}^{-1})+O_P\Big(\zeta_{Ld}^{1/2}\Big)\right]O_P(1)\\
&=&o_P((NT)^{-1/2}).
\end{eqnarray*}
Similarly, we can obtain that
\begin{eqnarray*}
s_2=\frac{1}{\sqrt{N}}\frac{1}{T^2}\sum_{s=1}^T\sum_{t=1}^T(\hat{F}_s-H^{\tau}F_s)^{\tau}
(\hat{F}_t-H^{\tau}F_t)\left(\frac{1}{\sqrt{N}}
\sum_{i=1}^NR_{is}\varepsilon_{it}\right)
\end{eqnarray*}
and
\begin{eqnarray*}
\|s_2\|&\le&
\frac{1}{\sqrt{N}}\left(\frac{1}{T}\sum_{t=1}^T\|\hat{F}_t-H^{\tau}F_t\|^{2}\right)
\left(\frac{1}{T^2}\sum_{t=1}^T\sum_{s=1}^T\left\|\frac{1}{\sqrt{N}}
\sum_{i=1}^NR_{is}\varepsilon_{it}\right\|^2\right)^{1/2}\\
&=&\frac{1}{\sqrt{N}}\left[O_P(\|\hat{\bm\gamma}-\widetilde{\bm\gamma}\|)
+O_P(\delta_{NT}^{-1})+O_P\Big(\zeta_{Ld}^{1/2}\Big)\right]^2O_P(1).
\end{eqnarray*}
For $s_3$,  by some simple calculations we have
\begin{eqnarray*}
s_3&=&\frac{1}{NT}\sum_{i=1}^N\frac{{\bm R}_i^{\tau}{\bm
F}}{T}HH^{\tau}(\hat{\bm F}H^{-1}-{\bm
F})^{\tau}{\bm\varepsilon}_i\\
&=&\frac{1}{NT}\sum_{i=1}^N\frac{{\bm R}_i^{\tau}{\bm
F}}{T}\left(\frac{{\bm F}^{\tau}{\bm F}}{T}\right)^{-1}(\hat{\bm
F}H^{-1}-{\bm
F})^{\tau}{\bm\varepsilon}_i\\
&&+\frac{1}{NT}\sum_{i=1}^N\frac{{\bm R}_i^{\tau}{\bm
F}}{T}\left[HH^{\tau}-\left(\frac{{\bm F}^{\tau}{\bm
F}}{T}\right)^{-1}\right](\hat{\bm F}H^{-1}-{\bm
F})^{\tau}{\bm\varepsilon}_i\\
&=:&s_{3.1}+s_{3.2}.
\end{eqnarray*}
Let $Q=HH^{\tau}-({\bm F}^{\tau}{\bm F}/T)^{-1}$. By Lemma
\ref{lemB3} (iii) and Lemma \ref{lemB2} (vi), we have
\begin{small}
\begin{eqnarray*}
s_{3.2}&=&\left(\frac{1}{NT}\sum_{i=1}^N\left[{\bm
\varepsilon}_i^{\tau}(\hat{\bm F}H^{-1}-{\bm
F})\otimes\Big(\frac{{\bm R}_i^{\tau}{\bm
F}}{T}\Big)\right]\right){\rm vec}(Q)\\
&=&\Big[(TN)^{-1/2}O_P(\|\hat{\bm\gamma}-\widetilde{\bm\gamma}\|) +
O_P(N^{-1})+N^{-1/2}O_P(\delta_{NT}^{-2})+N^{-1/2}O_P\Big(\zeta_{Ld}^{1/2}\Big)\Big]\\
&&\times\Big[O_P(\|\hat{\bm\gamma}-\widetilde{\bm\gamma}\|)+O_P(\delta_{NT}^{-2})+O_P\Big(\zeta_{Ld}^{1/2}\Big)\Big]\\
&=&N^{-1}O_P(\|\hat{\bm\gamma}-\widetilde{\bm\gamma}\|) +
N^{-1}O_P(\delta_{NT}^{-2})+
N^{-1/2}O_P(\delta_{NT}^{-4})+N^{-1}O_P\Big(\zeta_{Ld}^{1/2}\Big).
\end{eqnarray*}
\end{small}
Similar to the proof of $c_1$ in Lemma A.8 in Bai (2009s), we have
\begin{eqnarray*}
s_{3.1}=N^{-1}\psi_{NT}+(NT)^{-1/2}O_P(\|\hat{\bm\gamma}-\widetilde{\bm\gamma}\|)
+N^{-1/2}O_P(\delta_{NT}^{-2})+N^{-1/2}O_P\Big(\zeta_{Ld}^{1/2}\Big),
\end{eqnarray*}
where
\begin{eqnarray*}
\psi_{NT}=\frac{1}{N}\sum_{i=1}^N\sum_{j=1}^N\frac{{\bm
R}_i^{\tau}{\bm F}}{T}\left(\frac{{\bm F}^{\tau}{\bm
F}}{T}\right)^{-1}\left(\frac{\Lambda^{\tau}\Lambda}{N}\right)^{-1}\lambda_j
\left(\frac{1}{T}\sum_{t=1}^T\varepsilon_{it}\varepsilon_{jt}\right)=O_P(1).
\end{eqnarray*}
 For $s_4$,  note that $Q=HH^{\tau}-({\bm
F}^{\tau}{\bm F}/T)^{-1}$. Then,
\begin{eqnarray*}
s_4&=&\frac{1}{NT}\sum_{i=1}^N\left[{\bm\varepsilon}_i^{\tau}{\bm
F}\otimes\left(\frac{{\bm R}_i^{\tau}{\bm F}}{T}\right) \right]{\rm
vec}(Q)\\
&=&\frac{1}{\sqrt{NT}}\left(\frac{1}{\sqrt{NT}}\sum_{i=1}^N\sum_{t=1}^TF_t{\varepsilon}_{it}\otimes\left(\frac{{\bm
R}_i^{\tau}{\bm F}}{T}\right) \right){\rm vec}(Q)\\
&=&o_P(1),
\end{eqnarray*}
by the facts that  ${\rm
vec}(Q)=O_P(\|\hat{\bm\gamma}-\widetilde{\bm\gamma}\|)+O_P(\delta_{NT}^{-2})+O_P(\zeta_{Ld}^{1/2})$
and
$$\frac{1}{\sqrt{NT}}\sum_{i=1}^N\sum_{t=1}^TF_t{\varepsilon}_{it}\otimes\left(\frac{{\bm
R}_i^{\tau}{\bm F}}{T}\right)=O_P(1).$$
In summary, we have
\begin{eqnarray}\nonumber
&&\frac{1}{NT}\sum_{i=1}^N{\bm R}_i^{\tau}(M_{{\bm F}}-M_{\hat{\bm
F}}){\bm\varepsilon}_i\\ \nonumber
&=&N^{-1}\psi_{NT}+N^{-1/2}O_P(\|\hat{\bm\gamma}-\widetilde{\bm\gamma}\|^2)
+(NT)^{-1/2}O_P(\|\hat{\bm\gamma}-\widetilde{\bm\gamma}\|)\\\label{ApendB-6}
&&+N^{-1/2}O_P(\delta_{NT}^{-2})+N^{-1/2}O_P\Big(\zeta_{Ld}^{1/2}\Big).
\end{eqnarray}
Let ${\bm V}_i=N^{-1}\sum\limits_{j=1}^Na_{ij}{\bm R}_j$. Replacing
${\bm R}_i$ with ${\bm V}_i$, by the same argument we have
\begin{eqnarray}\nonumber
&&\frac{1}{NT}\sum_{i=1}^N{\bm V}_i^{\tau}(M_{{\bm F}}-M_{\hat{\bm
F}}){\bm\varepsilon}_i\\ \nonumber
&=&N^{-1}\psi_{NT}^*+N^{-1/2}O_P(\|\hat{\bm\gamma}-\widetilde{\bm\gamma}\|^2)
+(NT)^{-1/2}O_P(\|\hat{\bm\gamma}-\widetilde{\bm\gamma}\|)\\\label{ApendB-7}
&&+N^{-1/2}O_P(\delta_{NT}^{-2})+N^{-1/2}O_P\Big(\zeta_{Ld}^{1/2}\Big),
\end{eqnarray}
where $\psi_{NT}^*=O_P(1)$ is defined as
\begin{eqnarray*}
\psi_{NT}^*=-\frac{1}{N}\sum_{i=1}^N\sum_{j=1}^N\frac{{\bm
V}_i^{\tau}{\bm F}}{T}\left(\frac{{\bm F}^{\tau}{\bm
F}}{T}\right)^{-1}\left(\frac{\Lambda^{\tau}\Lambda}{N}\right)^{-1}\lambda_j
\left(\frac{1}{T}\sum_{t=1}^T\varepsilon_{it}\varepsilon_{jt}\right).
\end{eqnarray*}
Letting $\xi_{NT}^*=\psi_{NT}-\psi_{NT}^*$, and together with (\ref{ApendB-6})
and (\ref{ApendB-7}), we finish the proof of Lemma \ref{lemB5}.
\hfill$\Box$

{\lemma\label{lemB6} Assume that  assumption (A1)-- (A9) hold. We have
\begin{eqnarray*}
D(\hat{\bm F})^{-1}-D({\bm F})^{-1}=o_P(1).
\end{eqnarray*}
 }

\noindent{\bf Proof.}\ \  Similar to the proof of Lemma A.7 (ii) in \cite{Bai2009}, we can  show that
\begin{eqnarray}\label{LA6}
\|P_{\hat{\bm F}}-P_{\bm F}\|=O_P(\|\hat{\bm\gamma}-\tilde{\bm\gamma}\|)+O_P(\delta_{NT}^{-2}).
\end{eqnarray}
This leads to
\begin{eqnarray*}
&&D(\hat{\bm F})-D(\bm F)\\
&=&\frac{1}{NT}\sum_{i=1}^{N}{\bm R}_i^{\tau}({ M}_{\hat{\bm F}}-{ M}_{\bm F}){\bm
R}_{i} -\frac{1}{T}\Big[\frac{1}{N^2}\sum_{i=1}^N\sum_{j=1}^N{\bm
R}_i^{\tau}({ M}_{\hat{\bm F}}-{ M}_{\bm F}){\bm R}_ja_{ij}\Big]\\
&=&\frac{1}{NT}\sum_{i=1}^{N}{\bm R}_i^{\tau}(P_{\hat{\bm F}}-P_{\bm F}){\bm
R}_{i} -\frac{1}{T}\Big[\frac{1}{N^2}\sum_{i=1}^N\sum_{j=1}^N{\bm
R}_i^{\tau}(P_{\hat{\bm F}}-P_{\bm F}){\bm R}_ja_{ij}\Big].
\end{eqnarray*}
The norm of the first term in the above expression is bounded above by
$$\left\|\frac{1}{NT}\sum_{i=1}^{N}{\bm R}_i^{\tau}(P_{\hat{\bm F}}-P_{\bm F}){\bm
R}_{i}\right\|\leq\frac{1}{N}\sum_{i=1}^{N}\left(\frac{\| {\bm R}_i\|^{2}}{T}\right)\|P_{\hat{\bm F}}-P_{\bm F}\|=o_{P}(1).$$
Similarly,  the order of the second term  is also $o_{P}(1)$.
Noting that $[D(\hat{\bm F})+o_{P}(1)]^{-1}=D(\hat{\bm F})^{-1}+o_{P}(1)$,  we complete the proof of  Lemma \ref{lemB6}.
\hfill$\Box$

\bibliographystyle{dcu}

\bibliography{references}

\end{document}